\documentclass[aps,prd,amsmath,floats,floatfix, twocolumn,
superscriptaddress,nofootinbib,showpacs,longbibliography]{revtex4-2}
\usepackage[T1]{fontenc}
\usepackage[utf8]{inputenc}
\usepackage{lmodern}
\usepackage{verbatim}

\usepackage{epsfig}
\usepackage{url}
\usepackage{hyperref}
\usepackage{subfigure}
\usepackage{multirow}

\usepackage[normalem]{ulem} 

\usepackage{latexsym}
\usepackage{epsfig}
\usepackage{tabularx}
\usepackage{amsmath}
\usepackage{amssymb}
\usepackage{wasysym}
\usepackage{graphicx}

\usepackage{dcolumn}
\usepackage{verbatim}
\usepackage{enumerate,mdwlist}%lists
\usepackage[titletoc]{appendix}
\usepackage{amsfonts}
\usepackage{fancyvrb} % for "\Verb" macro
\usepackage{tikz} %diagrams
\usetikzlibrary{calc}

\usepackage[export]{adjustbox}

\usepackage[normalem]{ulem}%strikethrough text

\newcommand{\ba}{\begin{eqnarray}}
\newcommand{\ea}{\end{eqnarray}}
\newcommand{\be}{\begin{equation}}
\newcommand{\ee}{\end{equation}}

\newcommand{\bx}{{\bf x}}

\newcommand{\de}{{\rm d}}
\newcommand{\vect}[1]{\boldsymbol{#1}}
\newcommand*{\Cc}{\mathcal}%

\newcommand{\Mlz}{M_{Lz}}

\newcommand{\ycr}{y_{\rm cr}}
\newcommand{\Mhost}{M_{\rm host}}
\newcommand{\wow}{WOFs}

\definecolor{grey}{rgb}{0.4,0.4,0.4}
\definecolor{dullmagenta}{rgb}{0.4,0,0.4}
\definecolor{darkblue}{rgb}{0,0,0.4}
\definecolor{midblue}{rgb}{0,0,0.5}
\definecolor{midred}{rgb}{0.5,0,0}
\definecolor{orange}{rgb}{1,0.5,0}
\definecolor{lightbrown}{rgb}{0.75,0.5,0.25}
\definecolor{tan}{cmyk}{0.14,0.42,0.56,0}
\definecolor{djunglegreen}{cmyk}{0.99,0,0.52,0}
\definecolor{lightgreen}{rgb}{0,1,0}
\definecolor{olivegreen}{cmyk}{0.64,0,0.95,0.40}
\definecolor{midgreen}{rgb}{0.0,0.675,0.0}
\definecolor{darkgreen}{rgb}{0,0.5,0}
\definecolor{ceruleanblue}{rgb}{0.0, 0.2, 0.7}
\definecolor{burgundy}{rgb}{0.5, 0.0, 0.13}
\definecolor{hvred}{RGB}{186,12,47}
\definecolor{ste}{rgb}{0.01, 0.28, 1.0}
\hypersetup{
    colorlinks=true,
    linkcolor=ceruleanblue,
    filecolor=ceruleanblue,      
    urlcolor=midblue,
    citecolor=burgundy
}

\newcommand{\glow}{\texttt{GLoW}}
\usepackage[all]{xy} %array diagrams
\usepackage{amsfonts}

\makeatletter
\def\l@subsubsection#1#2{}
\makeatother

\begin{document} 

% \title{Weak Lensing, Wave Optics and Gravitational Waves}

% \title{Weak Lensing Diffractive Features in Gravitational Wave Signals}

\title{Signatures of dark and baryonic structures on weakly lensed gravitational waves}
% from supermassive black holes}

\author{Guilherme Brando}
\email{guilherme.brando@aei.mpg.de}
\affiliation{Max Planck Institute for Gravitational Physics (Albert Einstein Institute) \\
Am Mühlenberg 1, D-14476 Potsdam-Golm, Germany}

\author{Srashti Goyal}
\email{srashti.goyal@aei.mpg.de}
\affiliation{Max Planck Institute for Gravitational Physics (Albert Einstein Institute) \\
Am Mühlenberg 1, D-14476 Potsdam-Golm, Germany}

\author{Stefano Savastano}
\email{stefano.savastano@aei.mpg.de}
\affiliation{Max Planck Institute for Gravitational Physics (Albert Einstein Institute) \\
Am Mühlenberg 1, D-14476 Potsdam-Golm, Germany}

\author{Hector Villarrubia-Rojo}
\email{hectorvi@ucm.es}
\affiliation{Departamento de Física Teórica, Universidad Complutense de Madrid \\
28040 Madrid, Spain}
\affiliation{Max Planck Institute for Gravitational Physics (Albert Einstein Institute) \\
Am Mühlenberg 1, D-14476 Potsdam-Golm, Germany}

\author{Miguel Zumalac\'arregui}
\email{miguel.zumalacarregui@aei.mpg.de}
\affiliation{Max Planck Institute for Gravitational Physics (Albert Einstein Institute) \\
Am Mühlenberg 1, D-14476 Potsdam-Golm, Germany}

\begin{abstract}
Gravitational lensing offers a powerful tool for exploring the matter distribution in the Universe. 
Thanks to their low frequencies and phase coherence, gravitational waves (GWs) allow for the observation of novel wave-optics features (\wow{}) in lensing, inaccessible to electromagnetic signals. Combined with the existing accurate source models, lensed GWs can be used to infer the properties of gravitational lenses. The prospect is particularly compelling for space-borne detectors, where the high signal-to-noise ratio expected from massive black hole binary mergers allows \wow{} to be distinguished deep into the weak lensing regime, drastically increasing the detection probability. 
Here, we investigate in detail the capacity of the LISA mission to detect \wow{} caused by dark matter halos, galaxies and the supermassive black holes (SMBHs) within them. We estimate the total optical depth to be $\lambda_{\rm tot} \sim 6 \times 10 ^{-3}$ for the loudest binaries of total mass $M_{\rm BBH} \sim 10^6 M_{\odot}$, with the dominant contribution coming from SMBHs. We also find that \wow{} in low-mass binaries $M_{\rm BBH} \sim 10^4 M_{\odot}$ are more likely due to the central galaxies. Within our model of gravitational lenses, we predict $\mathcal{O}(0.1)-\mathcal{O}(1)$ weakly-lensed events to be detectable during the 5 years of LISA mission, depending on the source population models. 
We show that \wow{} signatures are very sensitive to the properties of dark-matter halos with $M_{\rm vir}\in (10^6-10^8)M_\odot$: increasing the compactness parameter by $\sim 3$ in that range raises the detection rate by $\sim 26$. Additionally, we show that collective effects from the complex inner halo structure can further enhance detectability. This suggests that lensed GWs in LISA will be an excellent probe of dark-matter theories, baryonic and halo sub-structures.
\end{abstract}
\date{\today}

\maketitle

{
  \hypersetup{hidelinks}
 \tableofcontents
}

\section{Introduction}

Gravitational waves (GWs) propagating through the universe undergo gravitational lensing: the deflection, delay and distortion of signals due to intervening gravitational fields~\cite{Schneider:1992}. These phenomena allow GWs to probe the matter distribution in the universe, complementing not only other GW insights into astrophysics~\cite{KAGRA:2021vkt}, but also insights on the matter distribution from observations in the electromagnetic (EM) spectrum. Lensed EM sources offer valuable insight into the matter distribution from cosmological and cluster scales~\cite{Planck:2018lbu,Bartelmann:2010fz,Clowe:2003tk} to (sub)halos~\cite{Vegetti:2012mc,Hezaveh:2016ltk,DiazRivero:2017xkd} and stellar or substellar objects~\cite{Vernardos2023microlensing,Mroz:2024mse,Niikura:2017zjd,Muller:2024pwn}. These analyses also provide valuable insights into fundamental physics, such as the nature of dark matter~\cite{Vegetti:2023mgp,Massey:2010hh}. 

The nature of GWs makes them highly complementary to EM observations when it comes to lensing. First, observable GWs are emitted at extremely low frequencies, allowing the emergence of wave-optic features \wow{} ~\cite{Takahashi:2003ix}, which are very unlikely to be observed on lensed EM sources. 
Second, GWs are coherent and negligibly absorbed by matter, which allows them to probe the densest regions of gravitational lenses~\cite{Tambalo:2022wlm}. 
And third, the existence of ab-initio theoretical models for GW emission offers the possibility to further characterize lensing~\cite{Cheung:2024ugg}.
These advantages have led to a surge in the interest to characterize GW lensing in the wave-optics regime and its potential to probe cosmic structures~\cite{Oguri:2020ldf,Cremonese:2021ahz,Choi:2021jqn,Basak:2021ten,Tambalo:2022plm,Savastano:2022jjv,Caliskan:2022hbu,Gao:2021sxw,Cheung:2020okf,Fairbairn:2022xln,Savastano:2023spl,Urrutia:2024pos,Caliskan:2023zqm}.
WOFs depend starkly on the ratio of the lens mass to the signal's wavelength. The frequency of detectable GW sources is considerably broad -- from nHz with pulsar-timing arrays, to a $\sim 100$ Hz for stellar-mass binary black holes probed by ground-based detectors-- allowing many mass scales of gravitational lenses to be probed~\cite{Antoniadis:2022pcn,InternationalPulsarTimingArray:2023mzf,IPTA:2023ero}.

Space-borne GW detectors the $\sim$ mHz frequency range are especially powerful to study WOFs. First, \wow{} at mHz frequency are associated with lensing by halos of $\sim 10^6-10^8M_\odot$, which are dominated by dark matter~\cite{Bullock:2017xww}. This makes them ideal systems to study dark matter properties, but also notoriously difficult to probe with EM observation. 
Second, mHz detectors will observe the merger of Massive Binary Black Holes (MBBHs) with an extremely high signal-to-noise-ratio (SNR), even for sources at very high redshift. The combination of loud and distant signals maximizes the detection prospects, facilitating the observation of \wow{} well into the weak lensing regime, in which a single image is formed~\cite{Savastano:2023spl}.
The recent adoption of the Laser Interferometer Space Antenna (LISA)~\cite{Colpi:2024xhw,LISACosmologyWorkingGroup:2022jok} by the European Space Agency, and other proposed facilities~\cite{Sesana:2019vho,Sedda:2019uro,Baibhav:2019rsa,Gong:2021gvw,Ajith:2024mie,Zwick:2024hag} further motivates the investigation of wave-optics lensing.

The prospect of detecting \wow{} was first estimated with a fixed impact parameter~\cite{Takahashi:2003ix}. In an exploratory work, some of us addressed the wave-optics single image regime in detail, including an estimate of detection probabilities for typical LISA sources~\cite{Savastano:2023spl}. Our analysis and others~\cite{Gao:2021sxw,Caliskan:2022hbu,Caliskan:2023zqm} found a substantial probability of detection, although strongly dependent on the detectability criterion employed. Crucially, all these analyses assumed lenses of all masses to follow the singular-isothermal sphere (SIS) profile, a first approximation to galactic-scale lenses~\cite{Oguri:2013mxl}. 
However, other studies using the Navarro-Frenk-White (NFW) profile, a more realistic description of dark-matter halos, found substantially lower probabilities~\cite{Fairbairn:2022xln,Guo:2022dre}. 
The important role played by the halo profile highlights the need for an in-depth investigation and more realistic modelling that includes not only dark-matter halos and subhalos, but also galaxies within them and other objects, such as central supermassive black holes (SMBHs). 

In this paper we investigate the probability of detecting \wow{} in the single image regime, considering realistic lenses and expected LISA sources. In Section~\ref{sec:setup} we introduce the profiles and abundances of lenses - dark-matter halos, subhalos, galaxies, and SMBHs - and the mathematical aspects of the wave-optics lensing in the single image regime. 
In Section~\ref{sec:probs_iso} we present the detection prospects for a typical MBBH source, showcasing the results for each independent lens type and the total. In Section~\ref{sec:lisa_cats} we extend the results to realistic populations of LISA sources. 
In Section~\ref{sec:comp_lenses} we present a new methodology to study lenses composed by multiple, nested objects. 
In Section~\ref{sec:discuss} we study the impact of dark-matter halo compactness on the detectability of \wow{}, discuss our overall predictions in the strong lensing regime and the assumption of axisymmetric lenses. We finally present our last remarks and conclude our findings in Section~\ref{sec:concl}.

\section{Framework}
\label{sec:setup}

We will begin by introducing the setup for our calculations. Sec.~\ref{sec:lenses} presents our assumptions on modelling dark-matter halos, galaxies and their supermassive black holes (SMBHs). Sec.~\ref{sec:gw_diffr} reviews the wave-optic lensing formalism and construction of lensed waveforms, Section~\ref{sec:detectability} discusses the detection criteria, fitting factor vs mismatch analysis and their effects on the lensing probabilities.

\subsection{Gravitational Lenses}\label{sec:lenses}

GWs propagating through the universe are lensed by gravitational fields of massive objects along their path. For objects whose lensing time delay is comparable to the GW period, wave-optics effects produce frequency-dependent distortions to the GW waveform. In the LISA frequency range, distortions are caused by lenses of masses $10^{6}-10^{9}~M_{\odot}$ for point masses, and until $10^{12}~M_{\odot}$ for extended lens objects.
Various dark and baryonic structures in this mass range are present in the universe, from dark-matter halos and subhalos to galaxies (central and satellites) and supermassive black holes (SMBHs) at the center of each galaxy.  Depending on their mass, a halo can host a central galaxy, likewise, subhalos can host satellite galaxies. Moreover, a SMBH is typically associated with each galaxy.

To account for their imprint on the GW signals, we will begin by discussing the abundance of these types of lenses, quantified by the \emph{optical depth}. We will then proceed to specify their characterization. In Table~\ref{tab:lenses_summary}, we summarize the density profiles, and normalization constants used in this work.
\subsubsection{Optical Depth}\label{sec:opt_depth}

We will work under the assumption that different objects contribute independently to the probability of detecting \wow{}. For a given source, one can compute the optical depth, $\lambda$, as an integrated cross-section ($\sigma$) of lenses along the line of sight. Therefore, for a given lens-source-observer system the \textit{differential optical depth} can be written as
\begin{align}\label{eq:diff_opt_depth_lens}
   \frac{\mathrm{d}\lambda_{L}}{\mathrm{d} \log M_{L}}=\int_{0}^{z_{S}} &\mathrm{d} z^{\prime} \frac{c\left(1+z^{\prime}\right)^{2}}{H(z^{\prime})}
    \sigma_L  \nonumber \\
   &\times\frac{\mathrm{d} n_{L}\left(z^{\prime}, M_{L} \right)}{\mathrm{d}\log M_{L}}\,,
\end{align}
where $L$ is the lens type label. The integral is taken over the lens redshift, $z^{\prime}$,  from $0$ to the source redshift, $z_S$. Moreover, we introduced the distribution of lenses per logarithmic interval of mass, $\mathrm{d} n_{L}/\mathrm{d}\log M_{L}$, which is a function of the lens mass, $M_{L}$. The cross-section of the lens, $\sigma_L$ depends on both the source and the lens properties; its computation will be addressed in Sec.~\ref{sec:cross_sec}. For each lens type, integrating its differential optical depth over the logarithmic mass range of interest returns its optical depth, $\lambda_L$.

The last term in Eq.~\eqref{eq:diff_opt_depth_lens}, characterizes the lens mass distribution. For halos, this is simply the halo mass function, computed either analytically~\cite{press_schechter,shteth_tormen}, or from fitting formulae calibrated with \emph{N}-body simulations~\cite{diemer20,despali16,watson13,angullo12,tinker08}, and is a function of the redshift. The distribution of subhalos, central galaxies and satellite galaxies, however, are functions of not only the redshift but also their host halo mass. Therefore, we can still compute the same integral for the differential optical depth as the one in Eq.~\eqref{eq:diff_opt_depth_lens}, but write the contribution of each extra component in terms of its host halo. For subhalos, this is simply captured by
\begin{align}\label{eq:sub_distr}
    \frac{\mathrm{d}n_{\rm sub}}{\mathrm{d}\log M_{\rm halo}} = \frac{\mathrm{d}n}{\mathrm{d}\log M_{\rm halo}}\int\mathrm{d}\log M_{\rm sub} \frac{\mathrm{d}N}{\mathrm{d}\log M_{\rm sub}},
\end{align}
 where $M_{\rm sub}$ is the virial mass of a subhalo hosted by the halo with virial mass $M_{\rm halo}$. The integral gives the cumulative number of subhalos inside a given host halo, and the halo mass function in front of the integral weights the probability of finding such host at a given redshift.

\begin{figure}[t!]
\includegraphics[width=0.99\columnwidth]{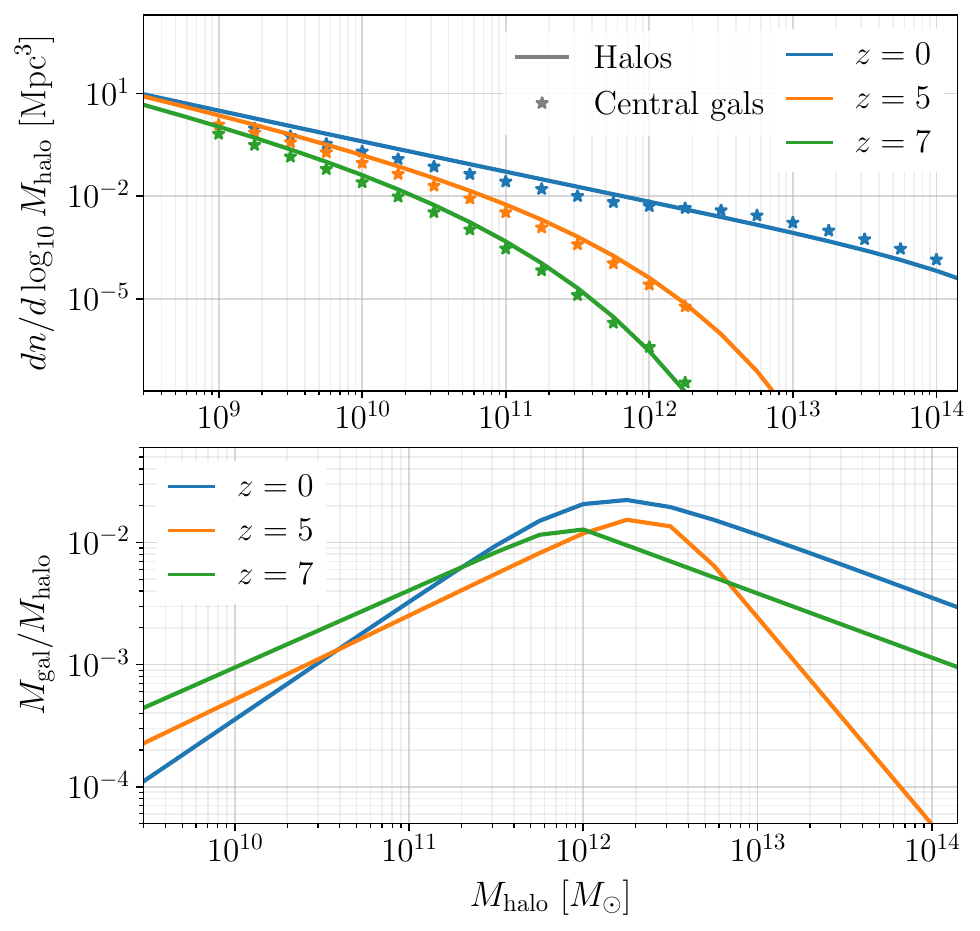}
\caption{\textbf{Top:} Halo and central galaxy abundance as a function of redshift and host halo mass. \textbf{Bottom:} Stellar to host mass relation for central galaxies, log-linearly extrapolated down to $10^{5}M_{\odot}$ as needed for this work (from Ref.~\cite{Behroozi:2019kql}).}
\label{fig:halo_gal}
\end{figure}

We can also write the abundance of central galaxies with stellar mass $M_{\rm gal}$ in terms of their host halo virial mass:
\begin{align}\label{eq:dngal_cen}
    \frac{\mathrm{d}n_{\rm gal, c}}{\mathrm{d}\log M_{\rm gal}} = \frac{\mathrm{d}\log M_{\rm halo}}{\mathrm{d}\log M_{\rm gal}}\frac{\mathrm{d}n_{\rm host}}{\mathrm{d}\log M_{\rm halo}},
\end{align}
where the first quantity on the right-hand side of Eq.~\eqref{eq:dngal_cen} is the stellar-to-host mass relation, shown in the bottom plot of Fig.~\ref{fig:halo_gal}. Similarly to Eq.~\ref{eq:sub_distr}, the satellite galaxy mass distribution is written as
\begin{align}\label{eq:dngal_sat}
    \frac{\mathrm{d}n_{\rm gal, sat}}{\mathrm{d}\log M_{\rm gal}} = \frac{\mathrm{d}n}{\mathrm{d}\log M_{\rm halo}}\int\mathrm{d}\log M_{\rm gal} \frac{\mathrm{d}\log M_{\rm sub}}{\mathrm{d}\log M_{\rm gal}}\frac{\mathrm{d}N}{\mathrm{d}\log M_{\rm sub}},
\end{align}
where the terms in the integral are, in order, the stellar-to-host mass relation for satellite galaxies and the subhalo distribution.  
The mass distribution of galaxies is modelled with the prescription outlined in Sec.~\ref{sec:lens_gals}. This ensures that central galaxies follow a Bernoulli distribution, and satellite galaxies a Poisson one.

SMBHs in this work are simply included in the lens distribution as proportional to the number of galaxies, assuming that $M_{\rm SMBH}=0.01 \times M_{\rm gal}$, following Ref.~\cite{Dosopoulou:2021kgs}. In Sec.~\ref{sec:prob_gal_bh} we will study how different values of this constant of proportionality affect our lensing estimates.

\begin{table*}[t]
    \centering
    \setlength{\tabcolsep}{10pt}
    \bgroup
        \def\arraystretch{1.5}% 
        \begin{tabular}{l c  c  c  c  c} \vspace*{0.1cm}
         Name                             & $\rho(r)$             & $\psi(x)$             & $\psi_0$                 & Parameters & Section               \\ \hline \hline
         PL                       & $\delta_D(r)$         & $\psi_0 \log(x)$                  & ${R_E^2}/{\xi_0^2}$          & -       & -               \\      \vspace*{0.1cm}
         SIS & ${\sigma_v^2}/({2\pi G r^2})$     & $\psi_0 x$                       & ${\sigma_v^2}/{(\xi_0 G \, \Sigma_{\rm cr})}$                       & $\sigma_{v}$          & \ref{app:lens_models} \\               \vspace*{0.1cm}
         NFW           & ${\rho_s}/\left[{(r/r_s) (1 + r / r_s)^2}\right]$ & $\frac{\psi_0}{2}\left[\ln^2\frac u2+(u^2-1)\mathcal{F}^2(u)\right]$                       & ${4 \kappa_s r_s^2}/{\xi_0^2} $ & $\rho_s, r_{s}$  & \ref{app:lens_models}  \\
         \hline \hline \vspace*{0.1cm}
    \end{tabular}
    \egroup
    \setlength{\tabcolsep}{6pt}
    \caption{Summary of lens models used in this work. Further details about these lenses (e.g.~normalisation, virial masses) and their phenomenology are discussed in the corresponding sections. The point lens is shown for comparison, with the Einstein radius defined as $R_E\equiv\sqrt{4 G d_{\rm eff} (1+z_L) M_{\rm PL}}$.
    }
    \label{tab:lenses_summary}
\end{table*}

\subsubsection{Dark matter}\label{sec:lens_dm}

We begin by modeling dark-matter halos and subhalos, which are smooth objects bound by gravity, where galaxies and other baryonic structures reside. The optical depth of dark matter in our Universe can be split as
\begin{equation}\label{eq:tau_dm}
    \lambda_{\rm DM} = \lambda_{\rm halo} + \lambda_{\rm sub},
\end{equation}
 where $\lambda_{\rm halo}$ is the contribution from the main halos and $\lambda_{\rm sub}$ accounts for the substructure, here dubbed as subhalos, that form within the halo. The $\lambda_{\rm halo}$ term contains the one-halo term of the halo model, which was already considered in previous works ~\cite{Savastano:2023spl,Choi:2021jqn,Caliskan:2023zqm}. In the halo model approach, the one-halo term refers to the concept that galaxies and substructures are formed within gravitationally bound regions in space comprised of dark matter; see Ref.~\cite{Asgari:2023mej} and references therein for a comprehensive review on the halo model. In this work, we will assume the abundance of halos per mass interval to be given by the Tinker halo mass function~\cite{tinker08} as implemented in the \texttt{colossus} package\footnote{\hyperlink{https://bdiemer.bitbucket.io/colossus/index.html}{https://bdiemer.bitbucket.io/colossus/index.html}}~\cite{diemer18}.

To include the contributions of subhalos, we need to account for their evolution history.
When two dark-matter halos merge, the less massive halo initially survives as a bound object, i.e. a subhalo, orbiting within the potential well of the more massive halo, i.e. the host halo. The subhalos are then subject to dynamical effects inside the host halo, stripping their mass and leading to either disruption or survival. Therefore, the subhalo mass function is distinguished into two definitions: the \emph{unevolved} subhalo mass function, and the \emph{evolved} subhalo mass function. The former describes the subhalo mass function of all objects accreted into the host halo including disrupted ones, whereas, the latter includes only those subhalos that have survived until an arbitrary time. Therefore, we are interested in the latter (evolved subhalo mass function) to investigate how substructure in host halos impacts the GW lensing probabilities. 

Following Ref.~\cite{Jiang:2014nsa}, the subhalo mass function is well described by a Schechter-like function,
\begin{align}\label{eq:dN_subs}
    \frac{\mathrm{d}N}{\mathrm{d}\log \left(M_{\rm sub}/M_{\rm vir}^{\rm host}\right)} = &p_{1}\left(p_{5}\frac{M_{\rm sub}}{M_{\rm vir}^{\rm host}}\right)^{p_{2}}\times \nonumber\\
    & \exp\left[-p_{2}\left(p_{5}\frac{M_{\rm sub}}{M_{\rm vir}^{\rm host}}\right)^{p_{4}}\right],
\end{align}
where $p_{1}, \dots, p_{5}$ are parameters of the model. These parameters depend on formation time, and different physical processes, and their value is computed by the companion code \cite{Jiang:2014nsa}. 

Since we want to study the impact of substructure in halos of different masses, we will assume that this subhalo mass function is valid independent of the host halo mass and use it to cover subhalos of mass down to $M_{\rm sub}=10^{5}~M_{\odot}$. Ref.~\cite{Jiang:2014nsa} studied and calibrated the above fit for host halos of mass spanning from $10^{11}~h^{-1}M_{\odot}-10^{15}h^{-1}M_{\odot}$\footnote{Note that this is the quoted value in the paper, and here $h$ is the reduced Hubble parameter. This is the only time we will quote masses in units of $h^{-1}M_{\odot}$, as the rest of this work we will use $M_{\odot}$ directly.}. However, we extend its application to smaller masses by a log-linear extrapolation of the original fit formula.

The density profile used in this work to model all dark matter objects is the NFW  profile ~\cite{Navarro:1995iw,Navarro:1996gj}, defined by the mass density 
\begin{equation}
    \rho(r)    =    \frac{\rho_s}{(r/r_s) (1 + r / r_s)^2}
    \;,
\end{equation}
where $r_{s}$ is the scale radius and $\rho_{s}$ is the density at this radius. This model has been found to fit well the spatial distribution of dark matter halos from \textit{N}-body simulations. An extensive discussion of this model is presented in App.~\ref{app:lens_models}, where we also present the associated normalization scale length used in our computations.

It is worth stressing that a fundamental parameter of the NFW profile is the halo concentration, $c_{\rm NFW}=r_{\rm vir}/r_{s}$, which is usually found by sampling the radial density distribution into discrete bins fitted by the NFW profile. This method has some shortcomings, as it is subjected to the choice of discrete bins, with small and large bins biasing the results in different ways. It is especially more prone to give incorrect estimations for low-mass halos, which are sparsely sampled. Analytical fits and descriptions for the NFW mass concentration relation, $c_{\rm NFW}-M$, also exist in the literature, such as~\cite{Diemer:2018vmz}. Later in this work, we will use the relation which is presented in~\cite{Ishiyama:2020vao}. Furthermore, we will consider that the same mass-concentration relation also holds for subhalos, which is an extrapolation of this fitting formula, since the concentration of dark matter substructure is known to be different than of halos~\cite{Moline:2021rza}.

The discussion regarding the NFW halo concentration is a topic of much debate. For example, in \cite{Wang:2019ftp}, the authors considered zoom-in simulations where dark matter is assumed to be a Weakly Interacting Massive Particle (WIMP) of mass $100~\text{Gev}$. This simulation has a large dynamical range, spanning $30$ orders of magnitude in mass and can resolve Earth-mass halos. Their results argue that at fixed mass the concentration of halos is tightly related to cosmology and the dark matter model. Therefore, the concentration model we have adopted does not encompass different theories of dark matter and is to be considered an agnostic prescription. Later in this work (cf. Sec.~\ref{sec:discuss_dm}), we will address the possibility of modifying this relation and investigate its impact on our lensing estimates.

%%%%%%%%%%%%%%%%%%%%%%%%%
%%------FIGURE-----------
%%%%%%%%%%%%%%%%%%%%%%%%%
%%%%%%%%%%%%%%%%%%%%%%%%%
%%------FIGURE-----------
%%%%%%%%%%%%%%%%%%%%%%%%%

%%%%%%%%%%%%%%%%%%%%%%%%%
%%------FIGURE-----------
%%%%%%%%%%%%%%%%%%%%%%%%%
\begin{figure}[t!]
\includegraphics[width=0.99\columnwidth]{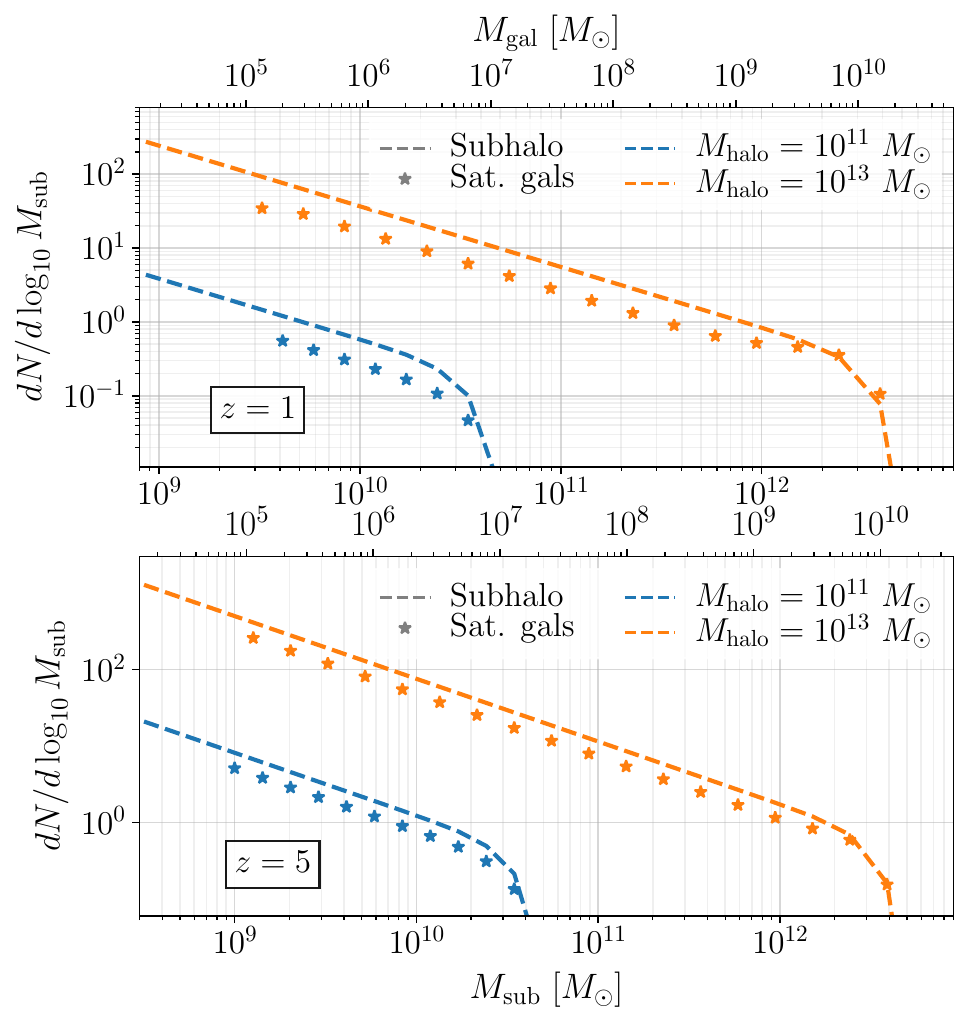}
\caption{Subhalo (dashed lines) and satellite galaxy (star) distribution as a function of the subhalo mass at two different redshifts, $z=1$ and $z=5$, for two different host halo mass, $M_{\rm halo}=10^{11}~M_{\odot}$ (blue) and $M_{\rm halo}=10^{13}~M_{\odot}$ (orange). The top axis of both plots shows the dependency of the stellar mass.}
\label{fig:sub_gal}
\end{figure}
%%%%%%%%%%%%%%%%%%%%%%%%%
%%------FIGURE-----------
%%%%%%%%%%%%%%%%%%%%%%%%%

\subsubsection{Galaxies}\label{sec:lens_gals} 

The gravitational potential of dark-matter halos and subhalos facilitates the formation of baryonic structures, such as galaxies. Because baryons interact, they can dissipate energy, achieving much higher densities than collisionless dark matter. 

We will consider both central galaxies (hosted by halos) and satellite galaxies (located in subhalos) and write the baryonic structures optical depth as the sum of both components,
\begin{equation}\label{eq:tau_lenses}
\lambda_{\rm baryons} = \lambda_{\rm gal, cent} + \lambda_{\rm gal, sat}.    
\end{equation}
We will model the abundance of galaxies living in a halo or subhalo by abundance matching~\cite{Wechsler:2018pic}, and we will assume a deterministic/functional correlation between the stellar mass and halo/subhalo mass, given by Ref.~\cite{Behroozi:2019kql}, using their publicly available data to generate the stellar to host mass relation~\footnote{\hyperlink{https://www.peterbehroozi.com/data.html}{https://www.peterbehroozi.com/data.html}}. 

Our modelling of the central galaxy abundance and relation to its dark-matter halo is shown in  Fig.~\ref{fig:halo_gal}.
The top panel shows the number density of halos and central galaxies as a function of the halo mass, for different redshifts. The bottom plot shows the stellar-mass to host-mass relation used in this work. Note that we are extrapolating log-linearly the fitting function down to $10^5 M_\odot$. This range of masses, while not calibrated with hydrodynamical simulations, is necessary to incorporate all objects that can produce \wow{} in LISA. Therefore, our results should be interpreted keeping in mind this uncertainty as well as the potential role of low-mass cutoffs.

Likewise, in Fig.~\ref{fig:sub_gal} we show the abundance of subhalos and satellite galaxies living in two different host halos at two redshifts. We model galaxies in this work by using the SIS lens profile:
\begin{equation}\label{eq:sis_rho}
  \rho(r)  
  =  
  \frac{\sigma_v^2}{2\pi G r^2}  
  \;,
\end{equation}
where $\sigma_{v}$ is the velocity dispersion of the galaxy. We assume the galaxy radius to be proportional to the virial radius of the host, i.e. $R_{\rm gal} \equiv f \times R_{\rm host}$, with typical values for $f$ being of the order of $f \sim 10^{-2}$. This has been shown by numerical simulations where $f$ was found to be nearly constant across many orders of halo mass and redshifts \cite{Karmakar_2023}. In Sec.~\ref{sec:prob_gal_bh}, we will explore the impact the value of $f$ on lensing probabilities. In App.~\ref{app:lens_models}, we expand on this lens model, presenting quantities derived from this lens profile in more detail. 

\subsubsection{Supermassive black holes}\label{sec:lens_smbhs}
The last lens type we will consider is SMBHs at the centre of each galaxy, modelled as point lenses. Studies have shown a scaling relation between the mass of the black hole and their residing halo~\cite{Ferrarese:2002ct}. A careful analysis was recently conducted comparing observations and the results of the IllustrisTNG~\cite{Pillepich:2017fcc, Nelson:2017cxy} and EAGLE~\cite{Schaye:2014tpa} cosmological simulations. However, in the present work we will pursue a more agnostic relation for the mass of SMBHs. Following reference~\cite{Dosopoulou:2021kgs}, we fix it to $M_{\rm SBMH}=0.01 \times M_{\rm gal}$, that is, one percent of the mass of the galaxy they inhabit. Later we will vary this ratio to test the impact of this assumption. 

We will not consider the possibility of wandering SMBHs, ejected from the galactic centers at galaxy mergers. Estimates suggest that typical galaxies host $\sim 5-15$ such objects with $M\gtrsim 10^6~M_{\odot}$~\cite{Bellovary:2010jr,2018ApJ...857L..22T}, however, we will consider only one SMBH per galaxy. The number scales linearly with halo mass, and each wandering black hole's mass remains close to its seed mass~\cite{Ricarte:2021gkf}. 

Finally, our total differential optical depth will then come from the contributions of individual lenses,
\begin{equation}\label{eq:tot_tau}
    \lambda_{\rm total} = \lambda_{\rm halos} + \lambda_{\rm subs.} + \lambda_{\rm gal, c} + \lambda_{\rm gal, sat} + \lambda_{\rm SMBH},
\end{equation}
with $\lambda_{\rm SMBH}$ is the sum of optical depths of SMBHs hosted by central and satellite galaxies.

\subsection{Wave-optics features}\label{sec:gw_diffr}

In this section, we review the basics of wave-optics propagation. We first introduce the lensing amplification factor and Green's function, which determines the lensed waveform in the frequency and time domains, respectively. Then, we discuss the \wow{} in the single-image weak lensing regime, compare different lens profiles, and conclude with the composite lens case.

\subsubsection{Lensing in the wave-optics regime}

The effect of lensing on the GW waveform is encoded in the frequency-dependent \emph{amplification factor}, 
\begin{equation}\label{eq:F_def}
    F(f)\equiv\frac{\tilde h(f)}{\tilde h_0(f)}\,,
\end{equation}
where $h(f), h_0(f)$ stand respectively for the unlensed and the lensed strain amplitude of the signal in the frequency domain. In the context of wave optics, this quantity is obtained by computing a diffraction integral \cite{Schneider:1992,Takahashi:2003ix,Tambalo:2022wlm}:
\begin{equation}\label{eq:amplification_F}
        F(w) = 
        \frac{w}{2\pi i}
        \int \de^2 \vect x 
        \exp\left(i w \phi(\vect x, \vect y)\right)
        \,.
\end{equation}
where the \emph{Fermat potential}, $\phi(\vect x, \vect y)$, is defined below.
The integration domain extends over the lens plane, with positions rescaled to a characteristic (arbitrary) normalization scale, $\xi_0$, so that $x$ is dimensionless. This choice sets the \emph{dimensionless frequency}:
\begin{equation}
    w\equiv 2 \xi_0^2 \frac{f}{d_{\rm eff}}\,=2\left(\frac{\xi_0}{r_F}\right)^2
\end{equation}
where the \emph{effective distance}, $d_{\rm eff}=\frac{D_L D_{LS}}{(1+z_L) D_S}$, depends on the lens, the source and lens-source angular diameter distances, indicated respectively as $D_{L}, D_{S}, D_{LS}$. In the second equality, we introduced the \emph{Fresnel length}, ${r_F \equiv \sqrt{{d_\mathrm{eff}(1+z_L)}/{(\pi f)}}}$, which determines the scale of onset of wave-optics effects. In this work, we adopt a normalization scale that is set by the definition of the \emph{redshifted effective lens mass},
\begin{equation}\label{eq:lensmass_def}
    \Mlz \equiv \frac{\xi_0^2}{4 G d_{\rm eff}}\,.
\end{equation}
With this choice, the dimensionless frequency becomes $w=8 \pi G \Mlz f$. Here and throughout the text we assume $c=1$.

The complex phase in the integral (\ref{eq:amplification_F}) depends on the Fermat potential 
\begin{equation}\label{eq:fermat}
    \phi(\vect x, \vect y) 
    \equiv
    \frac{1}{2}|\vect x - \vect y|^2- \psi(\vect x) + \phi_m(y)\,,
\end{equation}
consisting of the sum of the geometric and gravitational contributions to the time delay.
Here, $\boldsymbol{y}$ is the source position, projected onto the lens plane and normalized by $\xi_0$, and the lens model is specified through the \emph{lensing potential} $\psi(\vect x)$. The latter is sourced by the projected - onto the lens plane - matter density, $\rho(r)$, of the lens,  defined as
\begin{equation}\label{eq:Sigma}
    \Sigma(\boldsymbol{\xi}) = \int_{-\infty}^{+\infty}\mathrm{d}z \rho\left(\sqrt{z^{2} + \boldsymbol{\xi}^{2}}\right),
\end{equation}
where $z$ is the direction perpendicular to the lens plane, and the lensing potential is defined through the 2D Poisson equation, 
\begin{equation}\label{eq:psi_pot}
    \nabla_{\boldsymbol{x}}^{2}\psi(\boldsymbol{x}) =2 \frac{\Sigma(\xi_0 \vect x)}{ \Sigma_{\rm cr}} \equiv 2 \kappa(x).
\end{equation}
Here, ${\Sigma_{\rm cr} \equiv 1/(4 \pi G (1+z_L)d_{\rm eff})}$ is the critical surface density and the last equality defines the \emph{convergence} of the lens, $\kappa(x)$.
It is important to note that the Fermat potential is defined up to a constant, $\phi_m(\vect y)$, which we fix to set the minimum time delay to zero.

In the Geometric Optics (GO) limit, when $w\gg 1$, the diffraction integral is dominated by the stationary points of the Fermat potential,
$\nabla_{\vect x}\phi(\vect x_J, \vect y)=0$, and reduces to a discrete sum, where each term is associated to a GO image $J$:
\begin{equation}\label{eq:amp_factor_GO}
    F_{\rm GO}(w) 
    \simeq
    \sum_J \sqrt{|\mu_J|} \, e^{i w \phi_J- i\pi n_J}
    \,.
\end{equation}
Each image is distinguished by its proper apparent direction, denoted as $x_J$, along with the \emph{magnification factor} represented by $\mu^{-1} \equiv \det\left(\phi_{,ij}(\vect x_J)\right)$, and the \emph{time delay} denoted as $\phi_J \equiv \phi(\vect x_J,\vect y)$. Moreover, it acquires a \emph{Morse} phase correction, $n_J = 0$, $\pi/2$ or $\pi$, depending on whether $\vect x_J$ corresponds to a minimum, saddle point or maximum of $\phi$, respectively. The number of image produced depend on the lens type and the source position. For a given profile, the \emph{caustic curves} separate regions on the source plane corresponding to different image multiplicities.

Wave-optics predictions requires the full evaluation of the integral in Eq.~\eqref{eq:amplification_F}.
While closed-form analytical solutions are only available for the
point lens model, integrating the highly oscillatory complex exponential in \eqref{eq:amplification_F} is numerically challenging. One approach is to transform $F(w)$ to its time domain counterpart \cite{Ulmer:1994ij},
\begin{equation} 
 \label{eq:lensing_contour_time_integral}
  I(\tau)  \equiv \mathcal{F}\left(\frac{i F(w)}{w}\right) = \int \de^2 \vect x \, \delta\left( \phi(\vect x,\vect y) - \tau\right) \,,
\end{equation}
where $\mathcal{F}$ indicates the Fourier transform. The two-dimensional integral of $I(\tau)$ is performed by integrating over equal time-delay contours on the lens plane, and an inverse Fourier transform gives $F(w)$. Strategies to perform this integral efficiently for various lensing regimes were explored in \cite{Tambalo:2022plm, Savastano:2023spl}. In this work, the \glow{} code will be employed for these numerical evaluations.\footnote{\url{https://github.com/miguelzuma/GLoW_public}}

The derivative of $I(\tau)$ gives us the  \textit{lensing Green's function,} 
    \begin{equation}\label{eq:green_function_def}
        G(\tau) 
        \equiv 
        \frac{1}{2 \pi}
        \frac{\de}{\de \tau} I(\tau)
        \,.
    \end{equation}
The physical significance of these quantity is that its convolution with the unlensed waveform, $h_0$, yields the lensed waveform,
    \begin{equation}
        h(t) 
        = 
        \int_{-\infty}^{+\infty} \de t^{\prime} 
        \, G(t^\prime-t) h_0(t^\prime)
        \,,
    \end{equation}
where $t \equiv 4 G \Mlz \tau$. 
In the single image regime, $G(\tau)$ presents a singular contribution from the $I(\tau)$ discontinuity associated with the GO image at $\tau_I$. We can conveniently subtract it to get its regular part,
\begin{equation}\label{eq:green_function_split}
    \Cc G(\tau) 
    =    
    G(\tau) - \sqrt{|\mu|} \, \delta(\tau- \tau_I)
    \,,
\end{equation}
Hereafter, we will refer to $\Cc G(\tau)$ as the full Green's function. 
%%%%%%%%%%%%%%%%%%%%%%%%%
%%------FIGURE-----------
%%%%%%%%%%%%%%%%%%%%%%%%%
\begin{figure}[t]
    \centering
    \includegraphics[width=\columnwidth]{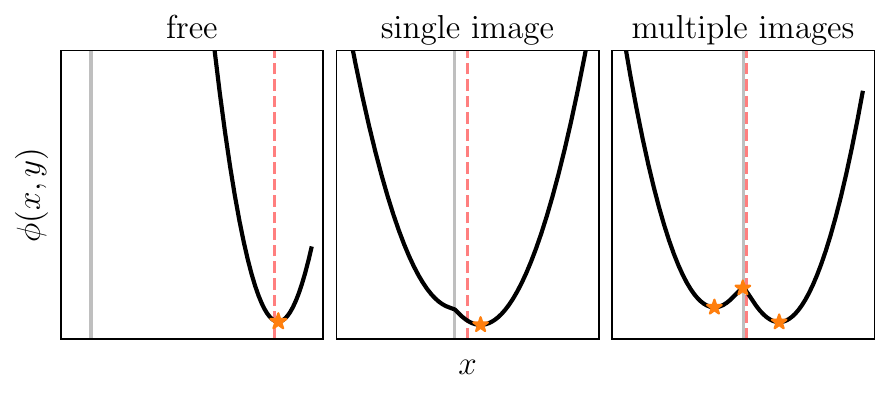}
    \caption{
    Illustrative diagram of the Fermat potential for a non-singular lens, e.g. NFW. Depending on the relative positions between the lens (solid grey) and the source (red dashed line), one or multiple GO images (orange stars) are produced. If the source-lens offset is large, as in the free case (left panel), the image almost overlaps with the source.  Their positions get displaced in the weak lensing regime (central panel), where the curve exhibits a central deformation that sources diffractive features.  In strong lensing (right panel), multiple GO images correspond to the Fermat potential's minimum, maximum, and saddle point.
    }
    \label{fig:diagram_fermat}
\end{figure} 
%%%%%%%%%%%%%%%%%%%%%%%%%
%%------FIGURE-----------
%%%%%%%%%%%%%%%%%%%%%%%%%
We are interested in the \textit{weak lensing} regime, in which the source's angular offset $y$ is large and lensing effects are subtle but potentially detectable. For an extended lens, a large enough offset will result in the formation of a single GO image. For a point lens, an additional image always forms, but its amplitude in the weak lensing regime decreases rapidly, as $\sqrt{|\mu|}\sim 1/y^2$. In weak lensing, the GO limit (Eq.~\ref{eq:amp_factor_GO}) only results in an overall rescaling and delay of the signal, which cannot be inferred from a single image. However, at finite $w$, diffraction effects can distort the waveform. In particular, beyond GO corrections to the amplification factor come from the region around the image and by non-differentiable points of the Fermat potential \cite{Takahashi:2004mc, Tambalo:2022plm}. For instance, the central cusp of lenses with a steep inner density profile produces such diffractive features. The phenomenology of this effect is extensively discussed in Ref. \cite{Savastano:2023spl} for the case of a singular isothermal sphere (SIS) model and its extension with a finite core. 

To illustrate this, we refer to the diagram in Fig. \ref{fig:diagram_fermat} showing the Fermat potential of a non-singular lens. For $y\to \infty$, the effects of lensing effects are negligible and the Fermat potential \eqref{eq:fermat} forms a parabola centered at its minimum, corresponding to the source position. In the strong lensing regime, at small $y$, the potential exhibits an additional saddle point and a local central maximum, giving rise to two lensing images. In the intermediate separation regime typical of weak lensing, the curve forms a deformed parabolic shape, resembling the strong lensing configuration but without the maximum and saddle point. This subtle deformation of the Fermat potential leads to potentially observable diffraction features. In comparison, the central maximum is replaced in the SIS model by a cusp with no image associated due to its steeper inner profile,  whereas the PL lens features a central divergence, which always ensures the formation of two images. In t at follows, we extend the analysis to consider more realistic axisymmetric lens profiles, each describing a specific gravitational lens, as discussed in Section~\ref{sec:lenses}.

\subsubsection{Isolated lens profiles}

The lens profiles we are interested in are the NFW, SIS and PL, as summarised in Table \ref{tab:lenses_summary}. In particular, dark matter halos are usually described by the NFW profile, while the SIS and PL can model galaxies and black holes, respectively; see Sec.~\ref{sec:lenses}. The NFW is akin to a broken power-law profile with finite density at the center. The PL has a Dirac delta function surface mass density, while the SIS has a singularity at the center and a caustic at $y=1$ where the image multiplicity changes. 
%%%%%%%%%%%%%%%%%%%%%%%%%
%%------FIGURE-----------
%%%%%%%%%%%%%%%%%%%%%%%%%
\begin{figure*}
    \centering
    \includegraphics[width=0.99\textwidth]{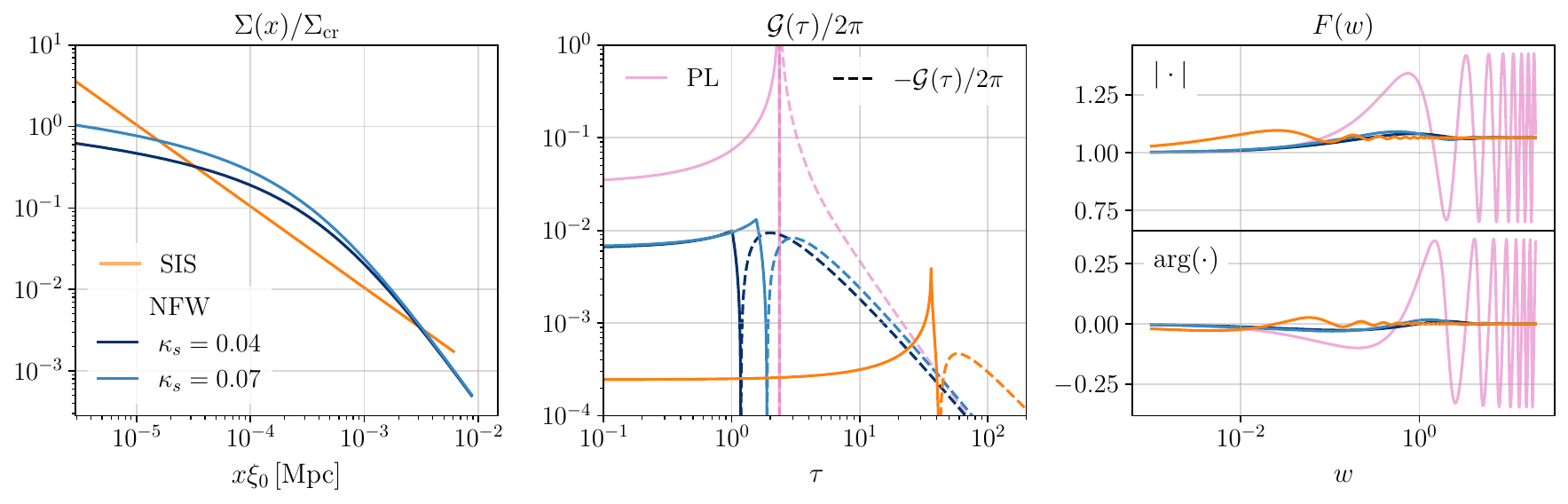}
    \caption{
    Comparison of different lens models: NFW varying $\kappa_s$ (blue shades), PL (magenta), SIS (orange). The left-hand side panel displays the surface density of the lens in physical units truncated at the virial radius, for an object with mass $10^9 M_\odot$. The central panel shows the time-domain lensing Green's function, $\Cc G(\tau)$. We compare weak lensing configurations sharing the time-domain amplification factor $I(\tau)$ at $\tau=0$, i.e. with the same primary image magnification. We set this value by fixing $y_{\rm SIS}, y_{\rm PL}=7.5, 1.12$ and $y_{\rm NFW} = 1.31 ,1.6$ for $\kappa_s=0.04, 0.07$. The right-hand side panels show the modulus and phase in the frequency domain, respectively. }
    \label{fig:diagram_ampl_fact}
\end{figure*}
%%%%%%%%%%%%%%%%%%%%%%%%%
%%------FIGURE-----------
%%%%%%%%%%%%%%%%%%%%%%%%%

Fig.~\ref{fig:diagram_ampl_fact} compares the weak lensing predictions of different lens profiles with projected density shown on the left panels (except for the point lens, which is a Dirac delta). The details on their implementation can be found in the App. \ref{app:lens_models}. 
We consider lenses with the same time-domain amplification factor value in the $\tau\rightarrow 0$ limit, i.e. matching $I(0)$. This corresponds to profiles with the same primary image magnification, $\mu\simeq I(0)/(2\pi)$. 

The middle panel shows the lensing Green's function, Eq.~\eqref{eq:green_function_split}. 
The point lens curve diverges in correspondence to the time delay of the secondary fainter image, which is produced even at large impact parameter. Similarly, the SIS lens features a sharp spike at the value of $\tau$ corresponding to the lens center (i.e. the cusp). The PL and the SIS have a significant broadband extension around their peaks, which is reflected in prominent \wow{}. 
The NFW profile is more complex and requires the specification of an additional parameter, the scale radius $r_s$; we show the curves for different convergence at the scale radius $\kappa_s=\kappa(x_s)$ values. In this case, the peak has finite magnitude and is narrower, depending on $\kappa_s$, with a larger value (denser inner region) producing a more pronounced feature. 

In the realistic configurations explored in this work, we will use profiles with $\kappa_s<0.3$  (see the discussion on the NFW concentration relation in App. \ref{app:lens_models} and Sec.~\ref{sec:discuss_dm} for the impact of higher concentration). This means the NFW feature will be generally weaker than the corresponding SIS with the same GO limit. The right panels of Fig.~\ref{fig:diagram_ampl_fact} show the frequency-domain result. At large $w$ we recover the GO limit; the amplification factor prediction matches for SIS and NFW, while the point lens' oscillates around this value due to the interference between the two images, in agreement with Eq.~\eqref{eq:amp_factor_GO}. Similar to the time domain, at low-$w$, the magnitude of the peaks depends on the central compactness of the lens. We stress that the time and frequency domain curves are plotted against the dimensionless variables $\tau$ and $w$. The conversion to physical quantities requires further specifications in the lens models. The NFW profile has a shallower central region than the SIS, which leads to less sharp diffraction effects in the weak lensing regime.

\subsubsection{Composite Lenses}\label{sec:comp_lens}

The profiles introduced above represent various types of cosmic structures. Alternatively to treating each lens as an isolated lens, we can instead describe the whole system as a \emph{composite} lens. The composite lens combines the individual lens potentials into a total one,
\begin{equation}\label{eq:comp_lens_psi}
      \psi^{\rm C}(\bx) \equiv \psi^{\rm host}(\bx) + \psi^{\rm gal}(\bx) + \psi^{\rm SMBH}(\bx).
\end{equation}
Here, we set the NFW host halo $\xi_{0}^{\rm host}$  as a common (for all profiles) normalization scale on the lens plane. As a consequence, the potential of each component scales with the ratio between the physical masses,
\begin{align}\label{eq:comp_lens_psi_0}
    \psi^{\rm gal}_0 \propto \sqrt{\frac{M_{\rm gal}^{4/3}}{M_{\rm host}}},\quad\textrm{and}\quad 
    \psi^{\rm SMBH}_0  \propto \sqrt{\frac{M_{\rm SMBH}}{M_{\rm host}}}\,,
    \;
\end{align}
where $M_{\rm host}$ is the mass of the host halo or subhalo. These scalings can be derived starting from the definitions in App.~\ref{app:lens_models} and the $\psi_0$ definitions in Table~\ref{tab:lenses_summary}. 

\begin{figure*}
    \centering
    \includegraphics[width=0.99\textwidth]{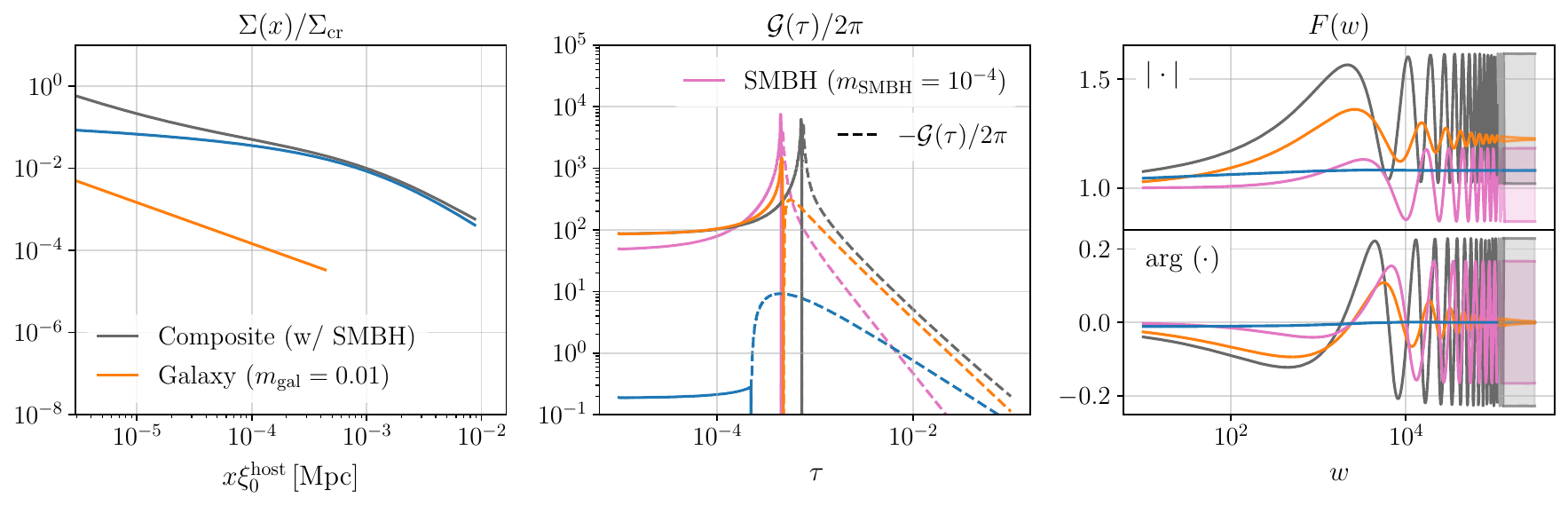}
    \caption{
    Comparison of isolated and composite lenses: Host (sub)halo (blue), SMBH (magenta), galaxy (orange), composite lens including the SMBH (grey). Here, we choose the impact parameter two times the SIS caustic radius, $y \times \xi_0^{\rm host} \simeq 5.8 \, {\rm pc}$, host mass $\Mhost=10^9 M_\odot$, and $m_{\rm SMBH}= 10^{-2}$ and $10^{-4}$. The source and lens redshifts are $z_S, z_L=5,2.5$. The left-hand side panel displays the surface density of the lens in physical units. The profiles are truncated at the virial (NFW host) or galactic radius (SIS galaxy with $f=0.05$). The central panel shows the lensing Green's function, $G(\tau)$. The right-hand side panels show the modulus and phase in the frequency domain, respectively. The shaded areas represent the curve's envelope in the high oscillatory GO region.
    }    
    \label{fig:ampl_fact_composite}
\end{figure*}

In Fig.~\ref{fig:ampl_fact_composite}, we present Green's functions and frequency domain amplification factors of the composite lens, along with its component contributions. We fix the component masses to $\Mhost=10^9 M_\odot$, $m_{\rm gal}\equiv M_{\rm gal}/M_{\rm host}= 10^{-2}$ and $m_{\rm SMBH}\equiv M_{\rm SMBH}/M_{\rm host}=10^{-4}$. The impact parameter is set to twice the caustic radius of the SIS profile that models the galaxy, $y \times \xi_0^{\rm host}\simeq 5.8 \, {\rm pc}$, ensuring weak lensing. Notably, the composite lens curve exhibits distinct behaviour compared to the individual components, emphasizing its unique characteristics. Specifically, the composite lens demonstrates the largest magnification in the GO limit  $w\rightarrow \infty$, and the largest magnitudes for the first wave-optics peak in $|F|$ and ${\rm arg}(F)$. From the plot we infer that combining lenses maximizes lensing signatures across all regimes. Furthermore, the SMBH (PL) contribution leads to the formation of a secondary image (saddle) in the composite lens. This additional feature influences the high-frequency behavior of the amplification factor, resulting in oscillations with a fixed amplitude, contrary to the expected weak lensing decay observed in the galaxy (SIS) and host (NFW) cases.
%%%%%%%%%%%%%%%%%%%%%%%%%
%%------FIGURE-----------
%%%%%%%%%%%%%%%%%%%%%%%%%
%%%%%%%%%%%%%%%%%%%%%%%%%
%%------FIGURE-----------
%%%%%%%%%%%%%%%%%%%%%%%%%
\subsection{\wow{}  Detectability}\label{sec:detectability}

In this section, we address the detectability of \wow{}. We first introduce our mismatch criterion to infer the lensing cross-section and discuss the results for various systems. In addition, we explore the role of parameter degeneracies between the source and the observer with a refined analysis.

\subsubsection{Detection Cross-section} \label{sec:cross_sec}

For axisymmetric lenses the detection cross-section $\sigma_L$ can be simply written as:
\begin{equation}
    \sigma_L = \pi \left(y_{\rm cr} \times \xi_{0}\right)^{2},
\end{equation}
where $y_{\rm cr}$ is the critical impact parameter up to which lensing signatures are detectable and $\xi_{0}$ is the characteristic length scale of the lens defined in Eq.~(\ref{eq:lensmass_def}).

As a criterion for detectability, we will use the mismatch between the lensed and unlensed waveform~\cite{Lindblom:2008cm}. This follows Ref.~\cite{Savastano:2023spl} and provides a simple criterion; the role of degeneracies with source parameters in the unlensed waveform will be discussed in Sec.~\ref{sec:ff} below. For two generic waveforms, denoted as $h_1$ and $h_2$, the mismatch is defined as:
\begin{equation}
\label{eq:def_mismatch}
    \mathcal{M} 
    \equiv 
    1-\frac{( h_1|h_2 )}{\sqrt{( h_1|h_1)}\sqrt{( h_2|h_2 )}}
    \,.
\end{equation}
Here, we used the definition of the noise-weighted inner product for two signals $h_1(t)$ and $h_2(t)$ with Fourier transforms $\tilde{h}_1(f)$ and $\tilde{h}_2(f)$:
\begin{equation}
    (h_1|h_2) \equiv 4\,\mathrm{Re} \int_{0}^{\infty} \frac{\de f}{S_n(f)} \tilde{h}_1(f)^* \tilde{h}_2(f)  \,,
\end{equation}
where $S_n(f)$ is the sky-averaged one-sided detector power spectral density. The SNR is defined in terms of this product as $\mathrm{SNR} \equiv \sqrt{(h|h)}$. Hereafter, $\mathcal{M}$ will stand for the mismatch between the lensed and unlensed waveform, unless otherwise specified. Following the discussion in \cite{Savastano:2023spl}, we consider \wow{} to be detectable if
\begin{equation}\label{eq:bayes_lindblom}
\mathcal{M}\times {\rm SNR}^2 \approx \ln \mathcal{B}  > 1  \,,
\end{equation}

We will focus on systems that can be observed by the future LISA detector. We consider equal-mass ratio, non-spinning compact binary coalescence with waveform described by the \texttt{IMRPhenomXHM} \cite{Garcia-Quiros:2020qpx} model in the \texttt{PyCBC} package \cite{alex_nitz_2023_7885796}. Our example source has total mass $M_{\rm BBH}=10^6\, M_\odot$ and is located at redshift $z_S=5$, while the lens is at $z_L=z_S/2$. The effect of the source's sky inclination and polarization relative to the detector is accounted for by averaging over the antenna pattern functions, as described in \cite{Robson:2018ifk}. With this procedure, we analyse the typical expected sources, which are neither optimally aligned nor close to a blind spot of the instrument. Finally, we neglect the detector's motion, since we work in the single-image regime and the SNR builds up mostly at the merger (see Sec.~IVA in Ref.~\cite{Tambalo:2022plm} or Ref.~\cite{Marsat:2020rtl}).

\begin{figure}[t]
    \centering
    \includegraphics[width=\columnwidth]{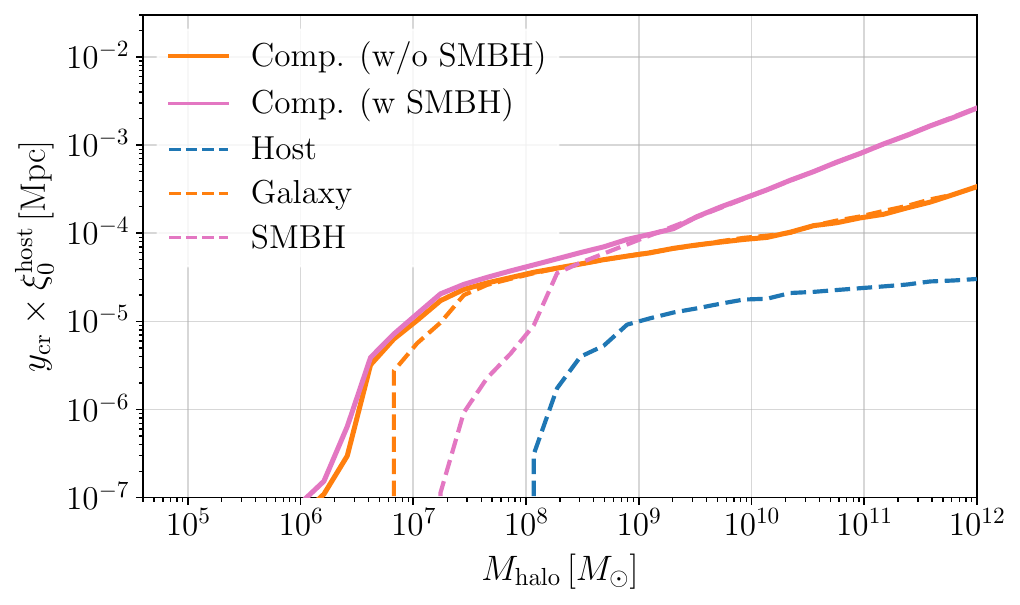}
    \caption{
    Critical impact parameter for detectability for a source with $M_{\rm BBH}=10^6\, M_\odot$, $z_S=5$ and $z_L=z_S/2$, as seen by LISA. The $\ycr{}$ values are presented in physical units and plotted against the host (sub)halo (NFW) mass. The comparison is made between the composite lens (solid line) and the individual component contributions (dotted lines), for $m_{\rm gal, SMBH}= 10^{-2}, 10^{-4}$ and $R_{\rm gal}/R_{\rm host}=0.05$. In the isolated case, the NFW host halo is shown in blue, the embedded SIS-like galaxy in orange, and the central galactic SMBHs in magenta. For composite lenses, the color corresponds to the dominant component.}
    \label{fig:y_crit_comparison}
\end{figure}

Fig.~\ref{fig:y_crit_comparison} shows the critical impact parameter, $\ycr{}$, in physical units, as a function of the host halo mass, represented by an NFW profile. As discussed above, the detectability for the host halo (dotted blue curve) is mainly driven by the convergence at the scale radius (itself dependent on both the concentration and mass). In particular, for $\Mhost<10^{8} M_\odot$, \wow{} from the lens centre become irrelevant, as $\kappa_s$ falls below $10^{-4}$ (cf. our assumptions on the model in App. \ref{app:lens_models}).  At larger host halo masses, the NFW profile becomes more compact and wave-optics effects can be detected up to a $\ycr\times \xi_0^{\rm host}\sim 10\,{\rm pc}$, while the curve drops as the signal enters in pure GO regime, for masses $\Mhost>10^{12} M_\odot$. 
In the Figure, a central embedded galaxy (dotted orange curve) with $m_{\rm gal}=10^{-2}$, produces a critical radius larger than the host halo at all mass values. As expected, the higher density of the SIS galaxy profile enhances \wow{}. Additionally, we observe that a central galactic SMBH with $m^{\rm SMBH}=  10^{-4}$, has an even larger detectable reach for $\Mhost{}>10^9 M_\odot$ (dotted pink curve). In this case, on top of wave diffraction, the interference between the two GO images (cf. Fig. \ref{fig:ampl_fact_composite}) can produce detectable beating patterns in the final signal for masses beyond $\Mhost=10^{12} M_\odot$, encompassing the range where the two images overlap and do not appear as separated events. 

In the composite lens case, $\ycr{}$ values mostly align with the dominant isolated component. However, for lighter host halos the composite lens can produce a higher $y_{\rm cr}$  than the isolated lens.  This is due to wave-optics effects being more relevant at low masses, as evident in Fig. \ref{fig:y_crit_comparison}. For example, for $M_{\rm halo}<10^{9} M_\odot$, the composite host+galaxy lens (solid orange curve) predicts a larger $\ycr{}$ than the isolated galaxy, which dominates the critical area over the host halo. This suggests that \wow{} are enhanced when the SIS galaxy is embedded in the NFW host halos and highlights the advantage of employing a composite lens over considering isolated lenses.
Compared to physical halo quantities, typical $\ycr{}$ values are more than 2 and 1 orders of magnitude smaller than the host halo and galactic virial radii, respectively, in the most optimistic case (host + galaxy + SMBH, solid pink curve).
%%%%%%%%%%%%%%%%%%%%%%%%%
%%------FIGURE-----------
%%%%%%%%%%%%%%%%%%%%%%%%%
%%%%%%%%%%%%%%%%%%%%%%%%%
%%------FIGURE-----------
%%%%%%%%%%%%%%%%%%%%%%%%%

\subsubsection{Role of parameter degeneracies}\label{sec:ff}

The mismatch criterion described above only includes a marginalization over the coalescence time and phase. Therefore, it does not account for potential degeneracies between the lens and the source parameters that may lower the significance of a detection. To address this degeneracy, we will compare the mismatch criterion with the \textit{fitting factor},
\begin{equation}\label{eq:fitting_factor_def}
1 - \mathcal{FF} \equiv \underset{\vec \theta_{\rm intr.}} {\text{min }} \mathcal{M} \approx \frac{\ln \mathcal{B}}{\text{SNR}^2}\,, 
\end{equation}
Here, the first equality defines $\mathcal{FF}$, and the mismatch is minimized over the source intrinsic parameters, $\vec \theta_{\rm intr.}=(M_{\rm BBH},q,\chi_{\rm eff})$, i.e. total mass, mass ratio and effective spin in addition to the coalescence time and phase. The second equality is an approximate relation for the Bayes factor~\cite{Basak:2021ten,cornish2011}. The critical impact parameter $y_{\rm cr}$ can be defined similarly to Eq.~\eqref{eq:bayes_lindblom} for a given significance, with $1-\mathcal{FF}\leq \mathcal{M}$ leading to more conservative results.

\begin{figure}[t]
\includegraphics[width=0.99\columnwidth]{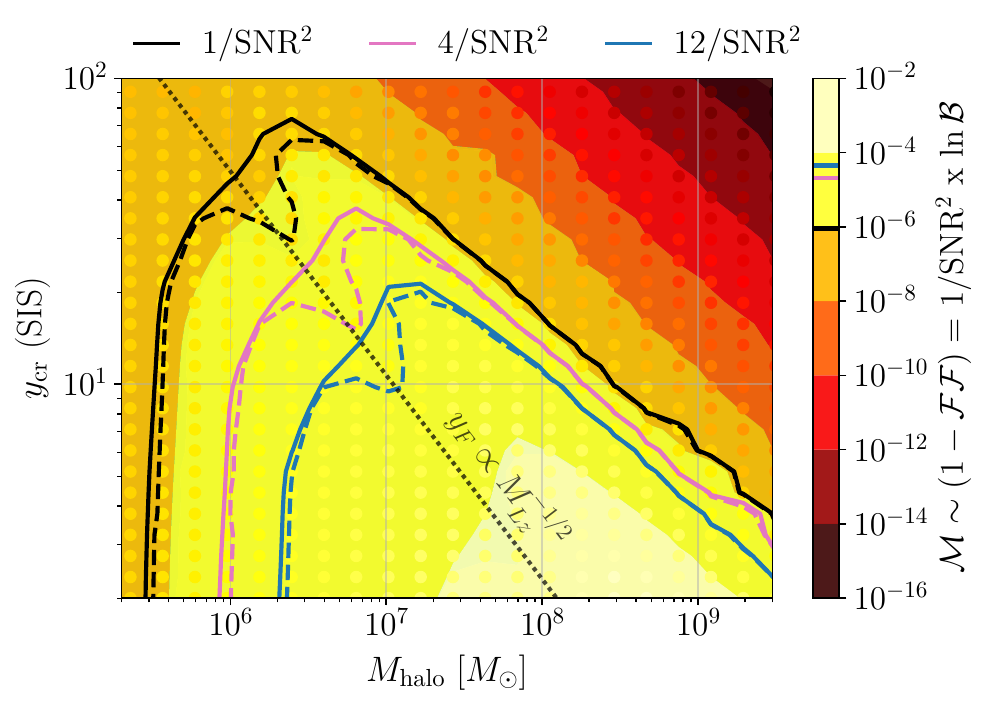}
\caption{Fitting factors (scatter) v/s the mismatch (contours) on a grid of dimensionless impact parameters and virial mass for an SIS lens and a source with $M_{\rm BBH}=10^6\, M_\odot$, $z_S=5$ and $z_L = 2.5$. The $y_{\rm cr}$ curves from fitting factors (dashed) are slightly smaller than the mismatch ones (solid) but follow a similar trend. The thresholds on Bayes factors $\ln \mathcal{B}$ correspond to approximately 1-3-5 $\sigma$ confidence levels. The dashed gray line represents the Fresnel length as a function of the virial mass. 
}
\label{fig:ff}
\end{figure}

While the fitting factor leads to more pessimistic detection prospects, the differences are small for most lens' masses. Fig. \ref{fig:ff} shows the mismatch values (filled contours) and fitting factors (scatter plot), both computed on a grid. Solid/dashed lines show the corresponding $y_{\rm cr}$ for different detection thresholds, $\ln \mathcal{B} =$ 1,4 and 12, which correspond to approximately 1, 3 and 5 $\sigma$ confidence levels, respectively (cf. Eq. 15 in \cite{pvalrev}). The main difference is a dent for lens masses slightly below the maximum of $y_{\rm cr}$, which aligns with the line defined by the Fresnel length $y_F\equiv r_F/\xi_0 \propto M_{ Lz}^{-1/2}$.

As expected, more conservative detection thresholds substantially decrease the critical impact parameter~\cite{Caliskan:2023zqm}. Because detection probabilities scale as $y_{\rm cr}^2$, the results will strongly depend on the detection thresholds. For simplicity, we quote most results for $\ln(\mathcal{B})=1$ (optimistic), but give some probabilities at different significance for comparison. More sophisticated methods like Fisher matrix or parameter estimation would be required for accurate results. We leave this for future work.
%%%%%%%%%%%%%%%%%%%%%%%%%
%%------FIGURE-----------
%%%%%%%%%%%%%%%%%%%%%%%%%
%%%%%%%%%%%%%%%%%%%%%%%%%
%%------FIGURE-----------
%%%%%%%%%%%%%%%%%%%%%%%%%

\section{Probabilities}\label{sec:probs_iso}

Based on the abovementioned methods, we now compute the probability of detecting \wow{}  associated with dark matter halos, galaxies, and SMBHs. We present the individual contributions to the optical depth and conclude by discussing the composite lens case.

\subsection{Dark-Matter Halos and Subhalos}\label{sec:prob_dm}

We start with the computation of the differential optical depth of halos and subhalos, for our example LISA source (cf. Sec.~\ref{sec:detectability}). Previous studies that modelled dark matter halos with an SIS profile found a peak in the differential optical depth for halos of low-mass $\sim 10^{6}~M_{\odot}$~\cite{Savastano:2023spl,Caliskan:2023zqm}. However, as we have discussed in Section~\ref{sec:lenses}, dark matter halos are better described by a Navarro-Frenk-White mass profile.
%%%%%%%%%%%%%%%%%%%%%%%%%
%%------FIGURE-----------
%%%%%%%%%%%%%%%%%%%%%%%%%
\begin{figure}[t]
\includegraphics[width=0.99\columnwidth]{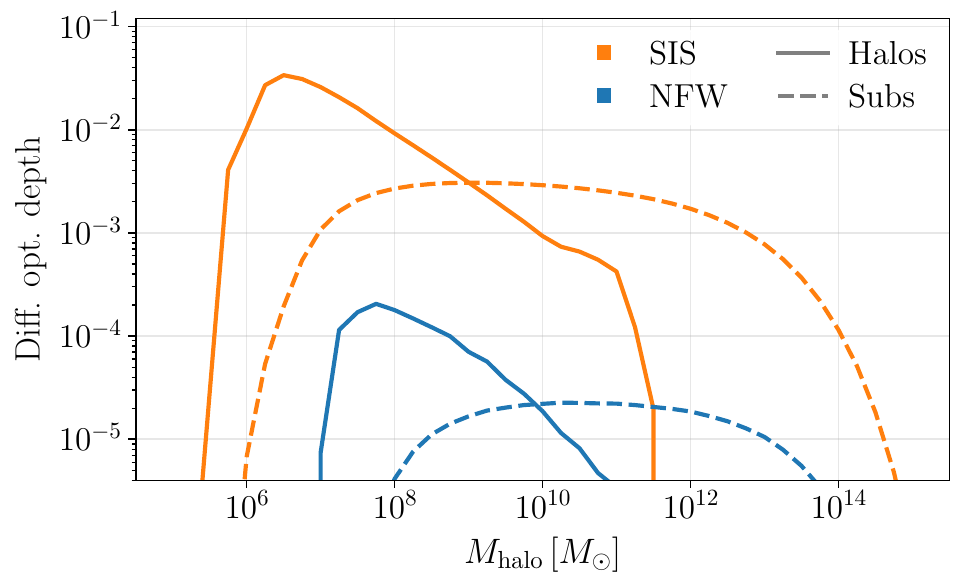}
\caption{Differential optical depth for dark matter only compact objects, halos (solid lines) and subhalos (dashed lines), following an SIS profile (orange) and NFW (blue). Here, we considered an equal-mass non-spinning BBH with $M_{\rm BBH}=10^6\, M_\odot$, $z_S=5$ and $z_L = 2.5$. }
\label{fig:halos_subs}
\end{figure}
%%%%%%%%%%%%%%%%%%%%%%%%%
%%------FIGURE-----------
%%%%%%%%%%%%%%%%%%%%%%%%%

Fig.~\ref{fig:halos_subs} shows a comparison between the differential optical depth for halos and subhalos, modeled as NFW and SIS, where the mass-concentration relation from~\cite{Ishiyama:2020vao} has been assumed (see Sec.~\ref{sec:lens_dm}). There is an overall reduction in the amplitude of approximately two orders of magnitude between the two profiles, both for halos and subhalos. The corresponding total optical depth, integrated over the virial mass of the host halos, is around $5.5\times 10^{-2}$ and $3.2\times 10^{-4}$, for the SIS and NFW curves, respectively. The latter value is consistent with Ref.~\cite{Urrutia:2023mtk}.
The reason for this difference is the much higher density at the center of SIS lenses, as it was anticipated in Fig.~\ref{fig:diagram_ampl_fact}. In particular, the SIS has a much stronger \wow{}  signature (peak in $\mathcal{G}(\tau)$ and oscillations in $|F(w)|$) than the NFW profile (for typical halo concentrations).

The subhalo predictions are shown as a function of the host halo virial mass (rather than the subhalo's actual mass), as discussed around Eq.~\eqref{eq:sub_distr}. Light subhalos that contribute to observable \wow{}  are more likely to occur in more massive host halos (see Fig.~\ref{fig:sub_gal}), which is why the differential optical depth extends towards high host masses, compared to the central (or isolated) halo contribution. We have only considered subhalos with mass $\gtrsim 10^6M_\odot$, as lower masses have a negligible chance of producing \wow{}  detectable by LISA. 
The optical depth for subhalos is $1\times 10^{-4}$ when modelled with an NFW profile, and $1.6\times 10^{-2}$ for the SIS. Thus, the optical depth for dark matter components, halos and subhalos, amounts to $4.1 \times 10^{-4}$ and $7\times 10^{-2}$ for the NFW and SIS, respectively.

Our NFW results strongly depend on the mass-concentration relation, and the impact of the concentration role on the detectability of \wow{} will be addressed in Section~\ref{sec:discuss_dm}. Furthermore, as Ref.~\cite{Heinze:2023ycq} recently pointed out, the subhalos density profile may differ from the NFW and exhibit a significant mismatch between the inner and outer logarithmic slopes, resulting in a steeper central region for a broad range of subhalos masses.

The observed decrease in the halo differential optical depth at high halo masses is due to the transition to GO. Structures with greater mass produce \wow{} observable at lower frequencies, below the LISA sensitivity.
However, these higher-mass halos contain a significant number of subhalos. Hence, the subhalo contribution extends further down to higher masses with a considerable amplitude, approximately one order of magnitude less than the maximum probability of the halos curve. Here the decrease at high masses is due to the scarcity of very massive host halos.
Moreover, for high halo mass, the differential optical depth for strong lensing (i.e. multiple images) becomes larger than that of \wow{} . In Sec.~\ref{sec:discuss_SL}, we will discuss the strong-lensing predictions within our models and compare them with results from simulations.

\subsection{Galaxies and Supermassive Black Holes}\label{sec:prob_gal_bh}

Let us now investigate how the addition of galaxies and their SMBHs impact the prospect of \wow{}  detection. As explained in Sec.~\ref{sec:lenses}, galaxies are modelled as SIS, as they are more compact than their host halos and subhalos. The additional compactness leads to more pronounced \wow{}, as can be seen in Fig.~\ref{fig:diagram_ampl_fact}, along with the different profiles.
Crucially, the mass of the galaxy depends on the redshift and host mass, see Figs.~\ref{fig:halo_gal} and \ref{fig:sub_gal}, which reduces the contribution of galaxies inside halos and subhalos (relative to modelling halos as SIS). The abundance of galaxies used to compute the cross-section (\ref{eq:diff_opt_depth_lens}) is given by Equations~(\ref{eq:dngal_cen}) and (\ref{eq:dngal_sat}). 
We also expect a SMBH at the center of most galaxies, and the amplification factor of a point lens is shown in Fig.~\ref{fig:diagram_ampl_fact}.

SMBHs are qualitatively different from extended lenses. Due to their compactness, they produce at least one additional GO image, whose amplitude decreases rapidly with the impact parameter in the weak lensing regime, with $\sqrt{|\mu|}\to y^{-2}$. 
The additional image can be understood from Fig.~\ref{fig:diagram_ampl_fact}: In the time domain, it appears as a logarithmic divergence in $\mathcal{G}(\tau)$. In the frequency domain, it corresponds to an oscillatory pattern that persists at all frequencies, instead of the characteristic damping of \wow{} (i.e. the interference between two GO terms in Eq.~\eqref{eq:amp_factor_GO}). Note that for typical SMBH masses, the time delay between images is shorter than the duration of the LISA signals: hence, they will be observed as superimposed waveforms. 
Unlike \wow{} , SMBHs detection is possible for masses $M_{\rm SMBH}\gg 1/(8\pi G f)$. This offers a qualitative difference with respect to the extended lenses considered in this work. 

\begin{figure}[t]
\includegraphics[width=0.99\columnwidth]{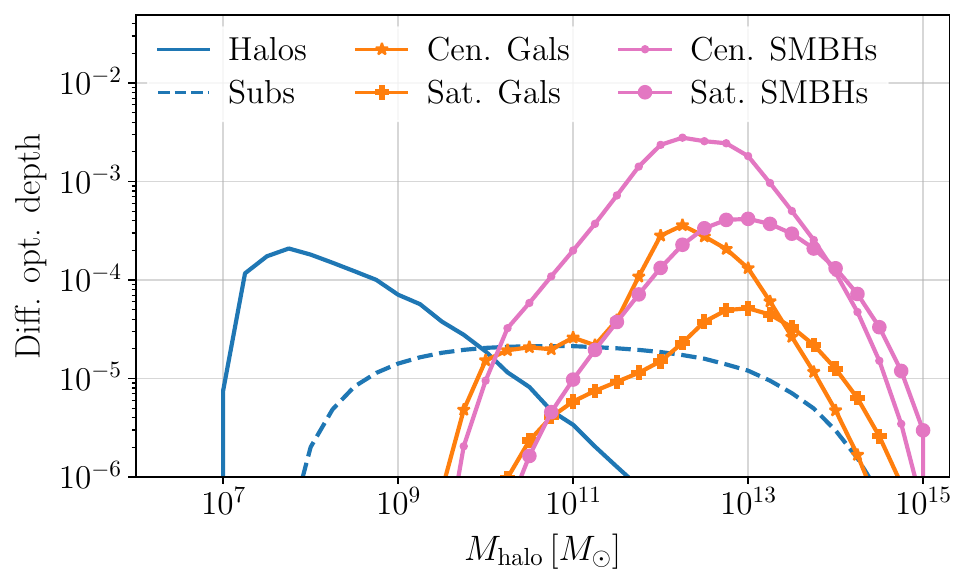}
\caption{Differential optical depth for dark matter only compact objects, halos (solid lines), subhalos (dashed lines), central galaxies (stars) and satellite galaxies (crosses), and SMBHs living in central galaxies (large dot) and satellite galaxies (smaller dot). Dark matter only objects are shown in blue, baryonic objects in orange and SMBHs in pink. The source is an equal-mass, quasicircular and non-spinning massive black hole merger with total mass $M_{\rm BBH}=10^{6}~M_{\odot}$ at redshift $z_{S}=5$. Here, we used $R_{{\rm gal}}/R_{\rm host}=0.02$.}
\label{fig:halos_subs_gals_bh}
\end{figure}

Fig.~\ref{fig:halos_subs_gals_bh} shows the differential optical depths associated to galaxies and SMBHs, both in host halos and subhalos. For this Figure we have fixed the galactic radius, discussed around Eq.~\eqref{eq:sis_rho}, to $R_{\rm gal}=0.02 \times R_{\rm host}$, and the mass of the SMBH to $M_{\rm SMBH}=0.01 \times M_{\rm gal}$. We summarize the differential optical depths for the example binary in Table~\ref{tab:halos_subs_gals_bhs}, where we see that the greatest contribution comes from SMBHs. We also compute optical depths for higher detection thresholds $\ln \mathcal{B} > $ 1-4-12, similar to  1-3-5 $\sigma$ detection of \wow{}. As seen in the different rows of the table, the optical depths reduce significantly ($\sim 10$ times) with higher detection thresholds.
%%%%%%%%%%%%%%%%%%%%%%%%%
%%------FIGURE-----------
%%%%%%%%%%%%%%%%%%%%%%%%%
%%%%%%%%%%%%%%%%%%%%%%%%%
%%------FIGURE-----------
%%%%%%%%%%%%%%%%%%%%%%%%%

%%%%%%%%--TABLE--%%%%%%%
\begin{table*}[t]
\centering
\renewcommand{\arraystretch}{1.}
%\begin{tabular}{| c | c | c | c | c | c | c |}
\begin{tabularx}{\linewidth}{X | X | X | X | X | X | X |  X }
\hline
$\ln \mathcal{B}$ &  Halos \hspace{1cm} (SIS)   & Subhalos \hspace{1cm} (SIS) & Central Galaxies  & Satellite Galaxies   & Central SMBHs   & Satellite SMBHs   & Sum \hspace{1cm}(SIS)  \\
\hline%\hline
 $  > 1 $ &$3.1 \times  10^{-4}$ ($5.3 \times 10^{-2}$) & $9.4 \times  10^{-5}$ ($1.6 \times 10^{-2}$)& $4.1 \times  10^{-4}$ & $9.5 \times  10^{-5}$ & $4.2 \times  10^{-3}$& $7.4 \times  10^{-4}$& $5.8 \times 10^{-3}$ ($6.9 \times 10^{-2}$) \\
 \hline%\hline
 $ > 4 $ &$2.3 \times 10^{-6}$ ($1.0 \times 10^{-2}$) & $5.4 \times 10^{-7}$ ($2.3 \times 10^{-3}$) & $3.8 \times 10^{-4}$ & $6.7 \times 10^{-5}$ & $1.0 \times 10^{-3}$& $1.7 \times 10^{-4}$& $1.6 \times 10^{-3}$ ($1.2 \times 10^{-2}$) \\
 \hline%\hline
 $ > 12 $ &$5.9 \times 10^{-8}$ ($4.4 \times 10^{-3}$) & $1.2 \times 10^{-8}$ ($6.4 \times 10^{-4}$) & $3.7 \times 10^{-4}$ & $6.3 \times 10^{-5}$ & $3.3 \times 10^{-4}$& $5.8 \times 10^{-5}$& $8.3 \times 10^{-4}$ \hspace{1cm} ($5 \times 10^{-3}$)
\\\hline
%\end{tabular}
\end{tabularx}
\caption{Individual and total lensing optical depths for a fiducial source with $M_{\rm BBH}=10^6\, M_\odot$, $z_S=5$. The dominant contribution comes from the SMBHs of the central galaxies, and for comparison, we also show the optical depths for halos with SIS lens profile assumption in parentheses. Rows correspond to different detection thresholds for \wow{} detection (similar to 1-3-5 $\sigma$ confidence levels). Higher thresholds reduce the lensing probabilities significantly. }
\label{tab:halos_subs_gals_bhs}
\end{table*} 
%%%%%%%%--TABLE--%%%%%%%

Despite there being fewer (satellite) galaxies than (sub)halos (see Fig.~\ref{fig:sub_gal}), we observe that galaxies exhibit a peak in the lensing probabilities at relatively high mass host halos (especially central galaxies). This is due to the stellar mass of such objects lying in the mass range of $M_{\rm gal}\sim 10^{6}-10^{8}~M_{\odot}$, where we have seen a peak in the signal of \wow{}. However, the total optical depth is not significantly impacted by the addition of galaxies as lens objects because of their scarcity and low mass.

\begin{figure}[t]
\includegraphics[width=0.99\columnwidth]{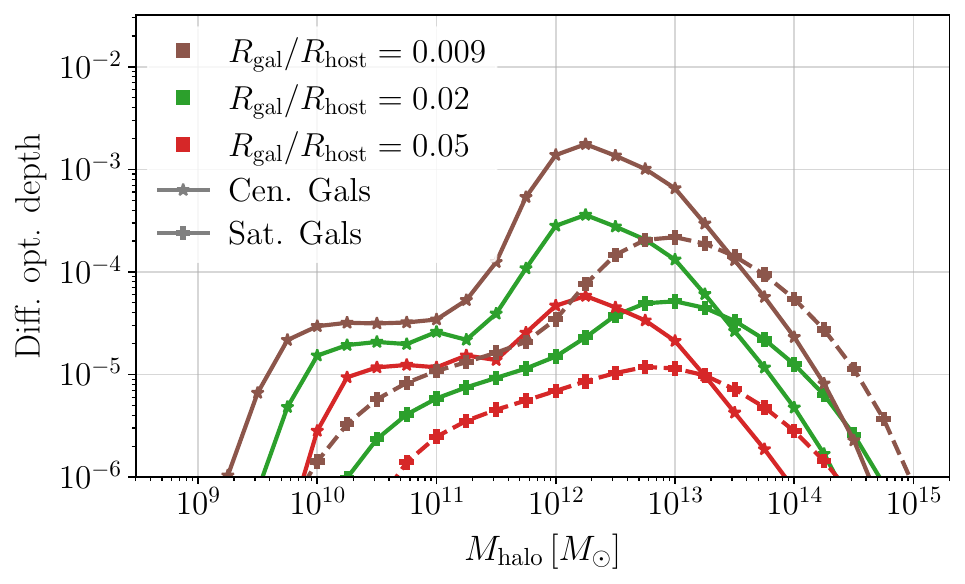}
\caption{Differential optical depth for central (stars) and satellite galaxies (crosses) varying the galaxy radius to be $0.009$ (brown), $0.02$ (green) and $0.05$ (red) of the size of the virial radius of its host. The source is an equal-mass, quasicircular and non-spinning massive black hole merger with total mass $M_{\rm BBH}=10^{6}~M_{\odot}$ at redshift $z_{S}=5$.}
\label{fig:gals_vary}
\end{figure}

Figure \ref{fig:gals_vary} illustrates how varying the galaxy radius relation, discussed in \ref{sec:lens_gals}, i.e. $R_{\rm gal}=f \times R_{\rm host}$, impacts the differential optical depth curves. We observe that if the galaxy radius is reduced to smaller values, for a fixed galaxy mass, there is an enhancement in detecting \wow{} by both satellite and central galaxies, with a greater chance of detection for the latter. For the cases considered in the Figure, the integrated optical depth for the combined contribution of galaxies, centrals and satellites, is $1\times 10^{-4}$, $5\times 10^{-4}$ and $2\times 10^{-3}$, for $f=0.05, \ 0.02$ and $0.009$, respectively. Therefore, a 5 times reduction of the galaxy size implies a 20 times enhancement in the probability of detection.

\begin{figure}[h]\includegraphics[width=0.99\columnwidth]{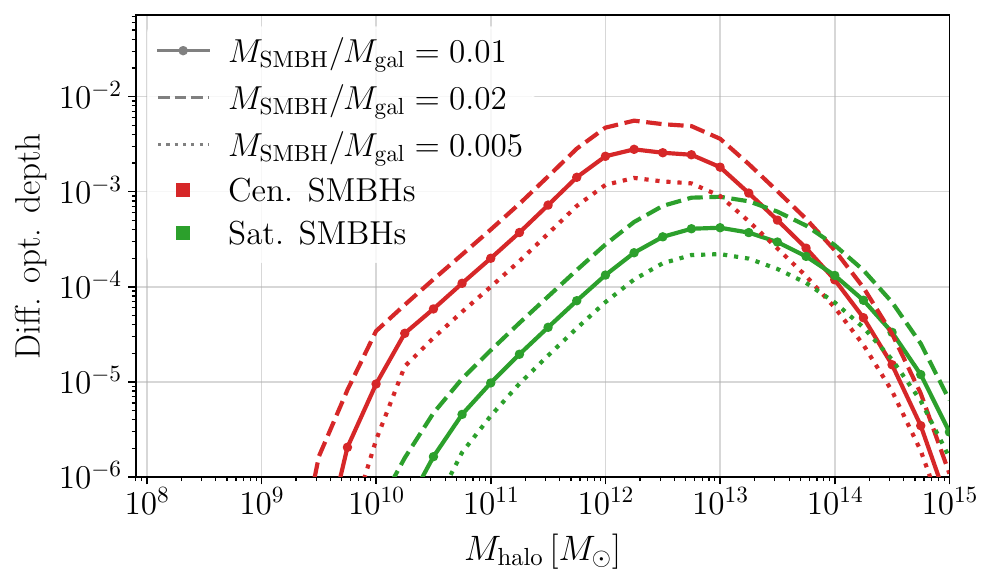}
\caption{Differential optical depth for SMBHs in central (red) and  (green) galaxies, varying the total mass of the SMBH with respect to their host galaxy by a factor of two.  The source is an equal-mass, quasicircular and non-spinning massive black hole merger with total mass $M_{\rm BBH}=10^{6}~M_{\odot}$ at redshift $z_{S}=5$.}
\label{fig:lens_prob_var_smbh}
\end{figure}

SMBHs produce the dominant contribution to the optical depth, primarily driven by the most massive objects (i.e. in massive host halos), for which either \wow{} or GO images are observable. 
To understand the impact of modifying the relation between the SMBHs and the host galaxies masses, in Fig.~\ref{fig:lens_prob_var_smbh} we show the differential optical depth curves for three different values of the ratio $M_{\rm SMBH}/M_{\rm gal}$. It is noticeable that when the fiducial value, $0.01$, is multiplied by a constant factor, the curves increase or decrease proportionally. Specifically, the figure examines cases where this mass ratio is doubled and halved, demonstrating the corresponding changes in the optical depth. The summed contribution from SMBHs living in central and satellite galaxies is $5\times 10^{-3}$, $1 \times 10^{-2}$ and $2.5\times 10^{-3}$, for $M_{\rm SMBH}/ M_{\rm gal}= 0.01, 0.02$ and $0.005$, respectively. The exact relation of SMBH and host mass is still an open question~\cite{smbhs_hydro}.

%%%%%%%%%%%%%%%%%%%%%%%%%
%%------FIGURE-----------
%%%%%%%%%%%%%%%%%%%%%%%%%
%%%%%%%%%%%%%%%%%%%%%%%%%
%%------FIGURE-----------
%%%%%%%%%%%%%%%%%%%%%%%%%

%%%%%%%%%%%%%%%%%%%%%%%%%
%%------FIGURE-----------
%%%%%%%%%%%%%%%%%%%%%%%%%
%%%%%%%%%%%%%%%%%%%%%%%%%
%%------FIGURE-----------
%%%%%%%%%%%%%%%%%%%%%%%%%

\subsection{Composite Lenses}\label{sec:comp_lenses}

%%%%%%%%%%%%%%%%%%%%%%%%%
%%------FIGURE-----------
%%%%%%%%%%%%%%%%%%%%%%%%%
\begin{figure}[h]
\includegraphics[width=0.99\columnwidth]{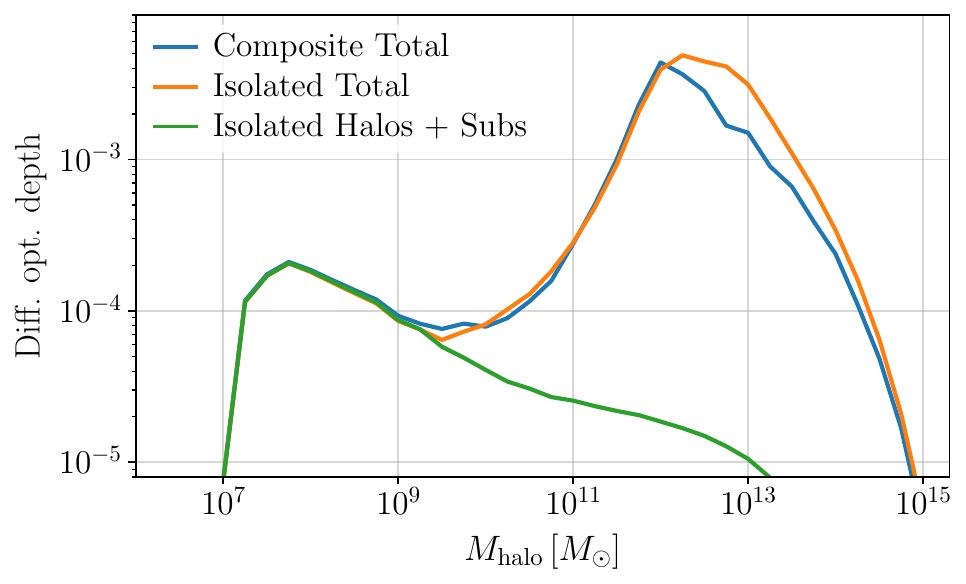}
\caption{Differential optical depth for isolated lenses, computed by summing each component of Fig.~\ref{fig:halos_subs_gals_bh}, from isolated dark matter only structure, computed by summing components of Fig.~\ref{fig:halos_subs} in blue, and the differential optical depth for a composite lens. The source is an equal-mass, quasicircular and non-spinning massive black hole merger with total mass $M_{\rm BBH}=10^{6}~M_{\odot}$ at redshift $z_{S}=5$. }
\label{fig:comp_lens_diff_opt}
\end{figure}
%%%%%%%%%%%%%%%%%%%%%%%%%
%%------FIGURE-----------
%%%%%%%%%%%%%%%%%%%%%%%%%

In previous subsections, we have presented differential optical depths computed by adding individually the contribution of isolated lenses. In such a scenario, the cross-section of each lens is assumed to not overlap with that of the other lenses. We will now employ the methodology of composite lenses presented in Section~\ref{sec:gw_diffr}, where the lensing potential of hosts (halos and subhalos) is combined with the contributions from galaxies and SMBHs.

Therefore, instead of taking Equation~(\ref{eq:tot_tau}) as the total optical depth, we can rewrite it as:
\begin{equation}
    \lambda_{\rm total} = \lambda_{\rm halo} + \lambda_{\rm sub},
\end{equation}
where the differential optical depth of each term on the right will still be computed by Eqs.~\eqref{eq:diff_opt_depth_lens}, with the distributions of halos and subhalos given by the usual halo mass function for the former, and Equation~(\ref{eq:dN_subs}) for the latter. Contributions from galaxies and black holes are, in this case, accounted directly in terms of the cross-section, which is now taken to have these structures embedded in an NFW profile. A comparison between the isolated lenses $\ycr$ curves and the composite lens counterparts was discussed when we presented Fig.~\ref{fig:y_crit_comparison}.

Fig.~\ref{fig:comp_lens_diff_opt} shows our results for composite lenses. The lines correspond to contributions from total composite lenses, the total for isolated lens analysis (Sec.~\ref{sec:prob_gal_bh}) and pure dark-matter halos and subhalos (Sec.~\ref{sec:prob_dm}). To understand the results of composite lenses, it is worth noting the behavior of the contributions from isolated halos and subhalos. At low masses, the composite lens results agree with the isolated dark-matter halos and subhalos, as no galaxies form in this regime. Deviations appear at larger masses because of galaxies and SMBHs populating the more massive objects.

Despite the different calculations, the contribution of composite lenses closely follows the results of the total contributions from isolated lenses at higher masses. This is expected, as pure wave-optics effects are imprinted by the same objects -- halos, subhalos, galaxies, and SMBHs. However, there are some subtle deviations, such as a slight difference at $M_{\rm halo}\sim 10^{10}~M_{\odot}$, where the composite lens prediction is slightly larger than its isolated counterpart. This manifests the differences in curves of isolated profiles in Fig.~\ref{fig:y_crit_comparison}, represented in dashed lines, and the composite lenses curves, solid lines. Notice that the level and the mass at which these differences manifest vary with lens redshift and stellar-to-host mass ratio.

Furthermore, the isolated lens result is slightly larger for $M_{\rm halo}>10^{13}M_\odot$. This feature is also expected, as precisely in this mass range where all contributions (dark-matter subhalos, galaxies, and SMBHs) are important, leading to a small number of overlapping cross sections, cf.~\ref{fig:halos_subs_gals_bh}. The reason for this is the relation between galaxy and host mass, Fig.~\ref{fig:halo_gal} and \ref{fig:sub_gal}. In the small mass range of Fig.~\ref{fig:halo_gal} we see a sharp decline in the stellar-to-host mass ratio, which implies that smaller mass halos do not host galaxies. The total optical depth of the composite lens is $4.5 \times 10^{-3}$, which is $\sim 30\%$ smaller compared to the isolated case. Overlapping of cross sections may not be the only reason for this small discrepancy, which may be due to the fact that to compute the differential optical depth for the composite lens cases we used a discrete grid of ratios between galaxy and host mass. Our grid ranges from $10^{-4}$ until $10^{-1}$, with four bins per decade and logarithmic spacing. This choice is motivated to contain the computational expenses in the computation of $y_{\rm cr}$.

\section{Detection prospects for LISA }\label{sec:lisa_cats}

After examining a typical MBBH merger (Sec.~\ref{sec:probs_iso}), let us consider the prospect of detection across a population of LISA sources. 
Following Ref.~\cite{Caliskan:2023zqm}, we consider a model-agnostic approach (Sec.~\ref{sec:pop_agnostic} and a model-dependent approach that reflects possible source formation channel (Sec.~\ref{sec:pop_catalogue}). While the model-dependent approach allows us to project the rates of observing wave-optics lensed MBBHs in LISA, the model-agnostic approach helps us to understand the dependence of \wow{} on source properties like masses, redshifts and SNR.  For the rest of this section, the total lensing probability is computed as the sum of individual (or isolated) lenses (Eq. \ref{eq:tot_tau}) following the methodology in section \ref{sec:probs_iso} and keeping the detection criteria as $\mathcal{M}$ $ \times $ SNR$^2 > 1 $. Note that in the following sections, we use probabilities and optical depth interchangeably as under small optical depth ($\lambda_L$) the lensing probability is equivalent to it.  Mathematically, if $\lambda_L \ll 1$ then,  $P(\lambda_L) = 1-e^{-\lambda_L} \approx \lambda_L$. 

\subsection{Model-agnostic}\label{sec:pop_agnostic}

Let us first examine the probability of detecting \wow{}  for LISA sources as a function of their mass and redshift. LISA is sensitive in detecting coalescing MBBHs up to very high redshifts with total masses in the range $10^4-10^8 M_{\odot}$. This analysis will help us understand \wow{}  detection prospects for sources not envisioned in the formation scenarios discussed below.

We compute the optical depths, $\lambda_{L}$, for various sources on a grid of detector-frame total mass $M_{\rm BBH}^{\rm D} \equiv M_{\rm BBH} (1 + z_{S})$ and source redshifts $z_S$. All sources are equal-mass, non-spinning and quasi-circular. The grid has 9 log-spaced bins in $M_{\rm BBH}^{\rm D}$  from $3\times 10^3$ to $3\times10^7$ and 21 bins in $z_{S}$ from $0.1$ to $20.1$.  Additionally, we put an SNR $> 8$ detection threshold, leading to about $\sim 150$ binaries.  For each of these binaries the lensing optical depth is calculated as per section \ref{sec:probs_iso} as the sum of individual isolated lenses. 

\begin{figure}[tbh]
\centering
\includegraphics[width=0.99\columnwidth]{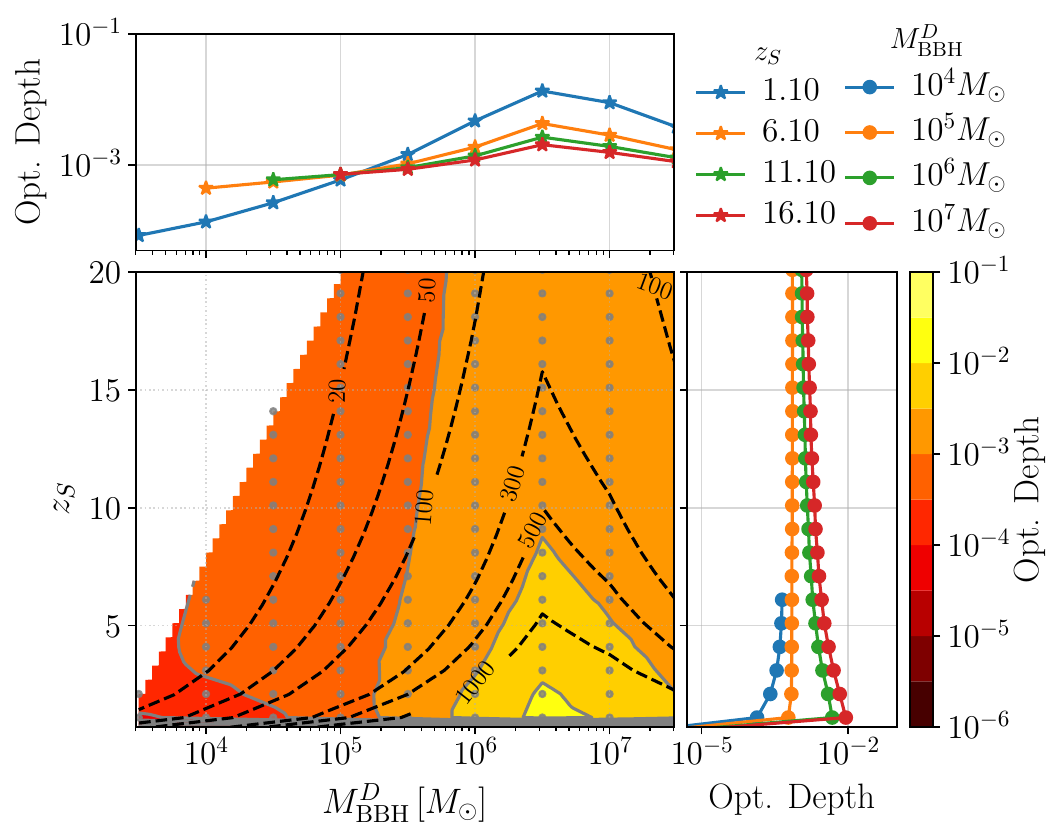}\\
(a) Total optical depth\\
\vspace{0.5cm}
\includegraphics[width=0.99\columnwidth]{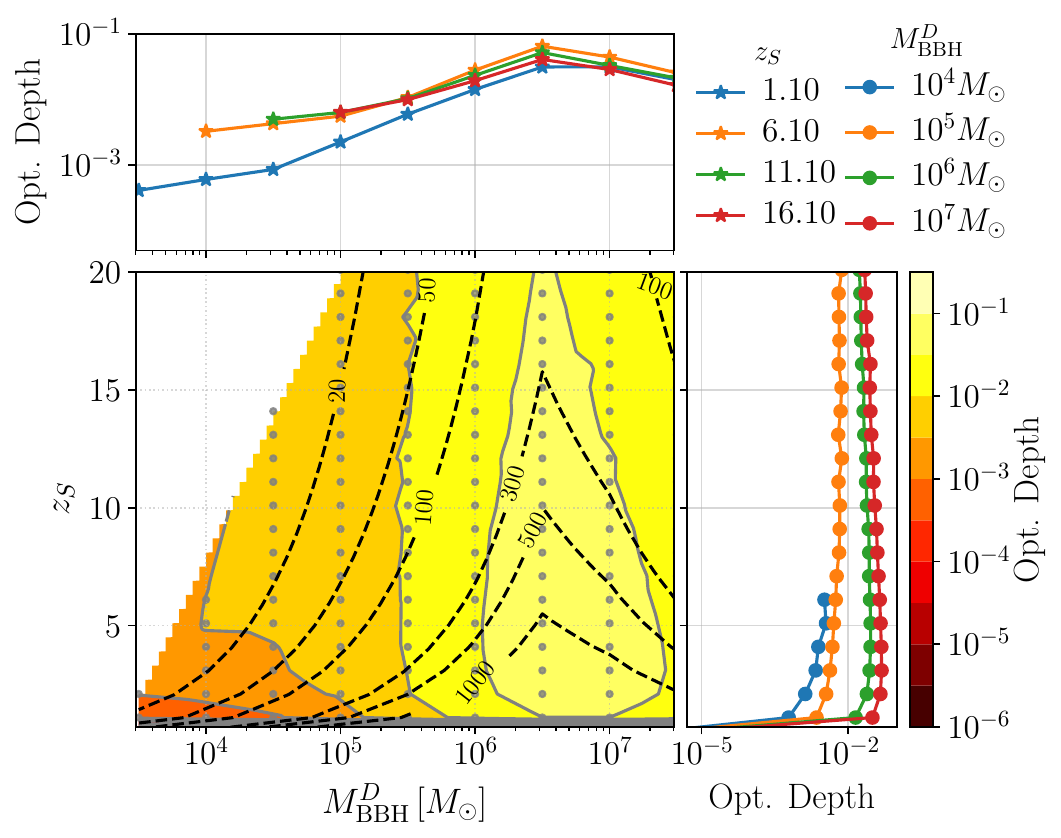}
(b) SIS halos and subhalos\\
\caption{(a) Total optical depth (computed as the sum of individual isolated lenses, see Eq.~\ref{eq:tot_tau}) for various LISA sources as a function of detector-frame total mass, $M_{\rm BBH}^{\rm D}\equiv (1+z_S) M_{\rm BBH}$, and source redshifts evaluated on a grid (gray dots). Black dashed lines are the SNR contours, and gray solid is lensing probabilities showing the positive correlation with SNRs. (b) Lensing optical depth considering only the halos and subhalos with SIS profiles instead. 
}
\label{fig:lens_prob_boosted}
\end{figure}

Fig. \ref{fig:lens_prob_boosted} shows the total lensing optical depth contours computed from interpolating values on the grid (gray points). $\lambda_{L}$ follows closely the equal-SNR contours (black dashed lines), indicating a good correlation between both quantities as expected from our discussion in Sec.~\ref{sec:detectability}.  
The optical depths decrease a little with larger source redshift (at fixed $M_{\rm BBH}^{\rm D}$). 
The dependence on the binary masses is non-monotonic, and influenced by both the detector sensitivity curve and the mass dependence of the halo mass function. This behaviour is also reproduced for SMBHs modelled as point mass lenses; however, each type of lens has a peculiar dependence on source properties.

In App.~\ref{app:lensingprob}, we show the dependence of the lensing optical depths for each type of lens on the MBBH parameters. For smaller mass binaries ($ M_{\rm BBH}^{\rm D} < 10^5 $) the lensing optical depth is dominated by central and satellite galaxies, otherwise by the central SMBHs consistent with Fig.~\ref{fig:halos_subs_gals_bh}.

Of all the binaries considered, the highest lensing probability is found to be $1.1 \times 10^{-2}$ for $M_{\rm BBH}^{\rm D} = 3.2 \times 10^6, z_S = 1.1$, having an SNR $= 6945$.  For comparison Fig. \ref{fig:lens_prob_boosted} also shows the contours of probabilities assuming halos and subhalos with SIS lens profile, which are also consistent with \cite{Caliskan:2023zqm} where they assume a Press–Schechter halo mass function and SIS lens profile for the dark matter halos.  As discussed in section \ref{sec:prob_dm}, the SIS profile is more concentrated than the NFW, and therefore, the probabilities rise by almost an order of magnitude. 

%%%%%%%%%%%%%%%%%%%%%%%%%
%%------FIGURE-----------
%%%%%%%%%%%%%%%%%%%%%%%%%
%%%%%%%%%%%%%%%%%%%%%%%%%
%%------FIGURE-----------
%%%%%%%%%%%%%%%%%%%%%%%%%

%%%%%%%%%%%%%%%%%%%%%%%%%
%%------FIGURE-----------
%%%%%%%%%%%%%%%%%%%%%%%%%
\begin{figure*}[t]
\centering
\includegraphics[width=\textwidth]{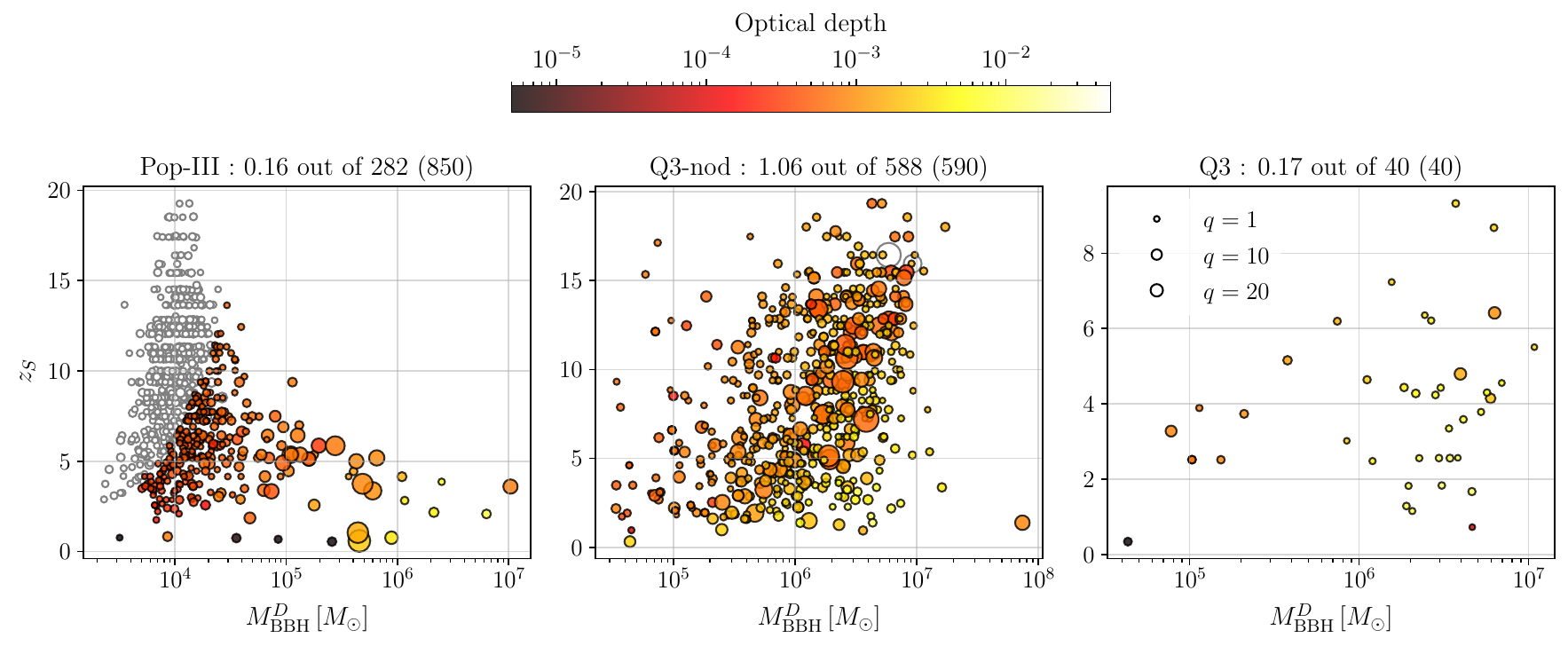}\\
(a) Simulated MBBH populations and their lensing optical depths in 5 years of LISA mission.\\
\vspace{0.5cm}
\includegraphics[width=\textwidth]{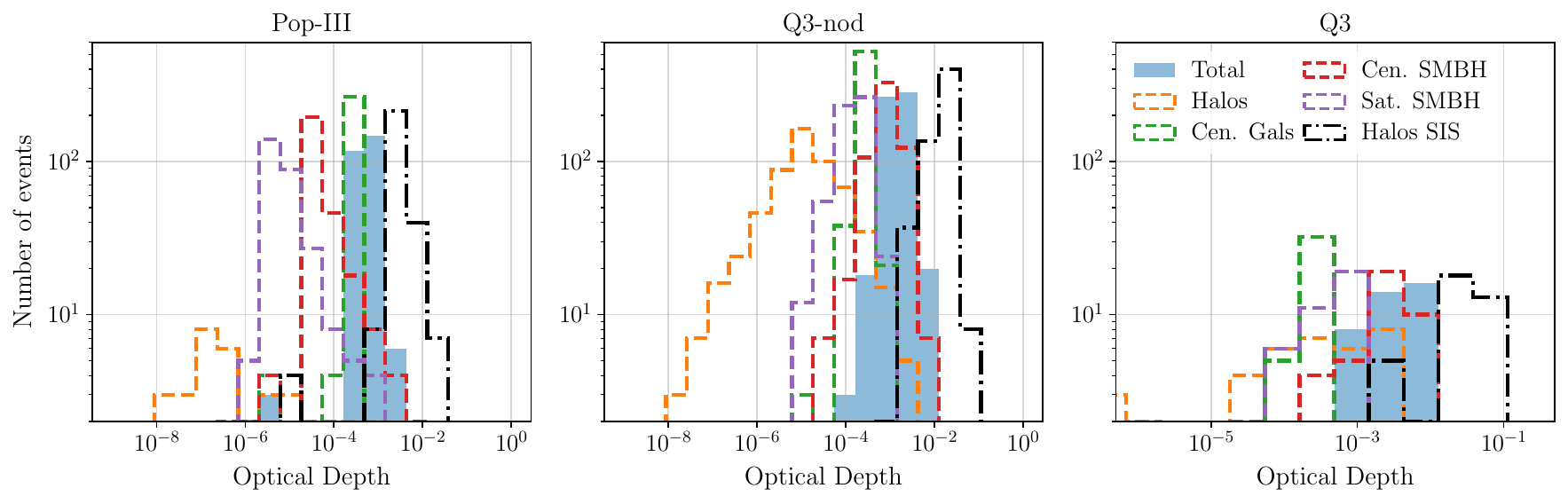}\\
(b) Distribution of the total and some of the individual optical depths.\\
\caption{ (a) Lensing optical depths for the detectable (having SNR $> 8$) MBBH populations for 5 years of LISA mission using various formation channels. Here, $M_{\rm BBH}^D\equiv (1+z_S) M_{\rm BBH}$ is the detector-frame total mass of the BBH. The mass ratio of binaries is proportional to the size of the markers and the undetectable binaries are the unfilled circles. Due to the low SNRs Population III formation channel (light seeds) has a smaller detection and lensing fraction than Q3 (heavy seeds). Q3-nod (heavy seeds without time delay) is the most optimistic scenario, with 588 detectable events out of 590 in 5 yrs, of which 1.06 events can be lensed. 
(b)  Distribution of total and individual optical depths for each type of MBBH populations. For Q3 and Q3-nod the optical depths are dominated by Central Galaxies SMBHs lenses, however for the Population III MBBHs the central galaxies contribute more. Under the assumption of halos and subhalos following the SIS profile, the lensing probabilities increase by an order of magnitude.
}  
\label{fig:lens_prob_LISA_pop}
\end{figure*}
%%%%%%%%%%%%%%%%%%%%%%%%%
%%------FIGURE-----------
%%%%%%%%%%%%%%%%%%%%%%%%%

%%%%%%%%%%%%%%%%%%%%%%%%%
%%------TABLE-----------
%%%%%%%%%%%%%%%%%%%%%%%%%
\begin{table*}[tbh]
\centering
\renewcommand{\arraystretch}{1.}
\begin{tabularx}{\linewidth}{X | X | X | X | X | X | X | X |  X | X }
\hline
 \multirow{2}{1.5cm}{Formation Channel} & \multicolumn{2}{c|}{Detection Rates} & \multicolumn{7}{c}{Average Optical Depth} \\ 
\cline{2-10}
    & \wow{}  \hspace{1cm} (SIS) & Unlensed  & Total \hspace{1cm} (SIS) & Halos \hspace{1cm} (SIS) & Subs \hspace{1cm} (SIS) & Cen. Gal. & Sat. Gal. & Cen. BHs & Sat. BHs\\
\hline
 Pop-III &  $0.16$ \hspace{1cm} ($1.2$) & 282  & $5.8 \times 10^{-4}$ ($4.2 \times 10^{-3}$) & $1.6 \times 10^{-5}$ ($3.8 \times 10^{-3}$) & $4.3 \times 10^{-6} $ ($5.9 \times 10^{-4}$) & $3.3 \times 10^{-4}$ & $5.4 \times 10^{-5}$& $1.5 \times 10^{-4}$& $2.5 \times 10^{-5}$ 
\\\hline
 Q3-nod &   $1.1$ \hspace{1cm} ($14.5$) & 588  & $ 1.8 \times 10^{-3}$ ($2.5 \times 10^{-2}$) & $7.5 \times 10^{-5}$ ($2 \times 10^{-2}$) & $2.1 \times 10^{-5} $ ($5.2 \times 10^{-3}$) & $3.6 \times 10^{-4}$ & $6.4 \times 10^{-5}$& $1.1 \times 10^{-3}$& $1.8 \times 10^{-4}$  
\\\hline
 Q3 &  $0.17$ \hspace{1cm} ($1.4$) & 40  & $4.3 \times 10^{-3}$ ($3.5 \times 10^{-2}$) & $5.9 \times 10^{-4} $ ($2.6 \times 10^{-2}$) & $1.7 \times 10^{-4} $ ($9.3 \times 10^{-3}$) & $2.8 \times 10^{-4}$ & $5.5 \times 10^{-5}$& $2.7 \times 10^{-3}$& $4.5 \times 10^{-4}$ 
\\\hline
%\end{tabular}
\end{tabularx}
\caption{Projected lensing rates for different formation channels in 5 yrs of LISA mission. The total lensing probabilities are computed as the individual isolated lenses and the contribution of each type of lens is shown. Optical depths assuming the halos and subhalos with SIS lens profiles are in parentheses for comparison.}
\label{tab:lisapop}
\end{table*} 
%%%%%%%%%%%%%%%%%%%%%%%%%
%%------TABLE-----------
%%%%%%%%%%%%%%%%%%%%%%%%%

\subsection{Astrophysical models} \label{sec:pop_catalogue}

The physics behind MBBHs formation is not yet well understood and several scenarios have been proposed. Ref~\cite{Klein:2015hvg} mentions the following formation channels for the MBBHs :  \textit{Population III} (light seeds), \textit{Q3} (heavy seeds) and \textit{Q3-nod} (heavy seeds without delay), which are briefly explained below. We point the reader to the original reference for more details.

\begin{enumerate}
    \item Light seeds or Population III : High mass, low metallicity primordial stars which undergo core-collapse supernovae to form massive black holes. They are called light-seeds by comparison, as they are in the mass range $10^2-10^4 M_{\odot}$ and form at a very high redshift ($\gtrsim 15$). The delay between binary formation and merger makes these MBBHs coalesce at lower $z$, thus being detectable by LISA.

    \item Heavy seeds (or Q3) black holes form directly from the collapse of proto-galactic disks at high redshifts ($\gtrsim 10$) and have a larger initial mass $\sim 10^5 M_{\odot}$. Two variants of this scenario are studied, with and without considering the time delay between formation and merger. The model without time delay Q3-nod, is the most optimistic population for LISA, giving the largest SNRs and total number of MBBH detections. However, after accounting for the time delay (Q3)  most of the heavy-seed binaries do not merge within the age of the universe, leading to low detection rates.
\end{enumerate}

We generate a population of MBBHs for each channel, using the sky-and-inclination angle averaged SNR threshold of 8. For simplicity, we assume all sources to be non-spinning. We find 282 out of 850, 588 out of 599, and 40 out of 40 sources detectable for the Population III, Q3-nod and Q3 formation channels, respectively, in 5 years of LISA mission. 
We compute the lensing probabilities for each detectable source, as described in sec. \ref{sec:probs_iso} by interpolating the results from Sec. \ref{sec:pop_agnostic} which hold for equal-mass binaries. Asymmetric binaries (mass ratio $q > 1$ ) have lower SNRs compared to symmetric binaries of the same total mass due to their low signal duration. We account for this change in SNR for sources with $q > 5 $ during the computation of probabilities but neglect the higher-order modes in the mismatch computations. The total \wow{}  optical depths $\lambda_{\rm total}$ (Eq. \ref{eq:tot_tau}) are shown in Fig.~\ref{fig:lens_prob_LISA_pop}. Notice that larger mass ratio binaries have lower SNRs and hence lower $\lambda_{\rm total}$. 

The overall \wow{}  rates, along with contribution from individual lens types, are outlined in Table \ref{tab:lisapop}.  As expected, the most optimistic formation channel Q3-nod (heavy seeds without delay) has the highest lensing rates of 1.06 out of 588 whereas, Population III (light seeds) has the lowest rates of about 0.16 lensed out of 282 events due to the low SNRs. Q3 channel has relatively low rates, 0.17 lensed out of 40 although it has the highest lensing probability, $4.3 \times 10^{-3}$ of detectable MBBHs. We also see in the lower panel of Fig.~\ref{fig:lens_prob_LISA_pop}  and Table \ref{tab:lisapop} that the dominant contribution comes from SMBHs (of both central and satellite galaxies), main halos, and central galaxies.

In contrast, if one considers only lensing by halos and subhalos following the SIS lens profile, the rates increase by an order of magnitude, with the highest being 14.5 for the Q3-nod formation channel. Our rates for the dark matter modelled as SIS lenses are consistent with that of \cite{Caliskan:2023zqm} (where they assume lenses following Press–Schechter halo mass function) even though the methods are different. As discussed in the previous sections the lensing probabilities and hence the rates will reduce if one increases the detection thresholds on the Bayes factors. 

Recent observation of GW background from inspiraling SMBHs in the nano-Hz regime by pulsar timing arrays suggests a median of $\mathcal{O}(75)$ SMBHs binaries per year detected by LISA in the mass range ${M}_{\rm BBH} \in  [10^6,10^7] M_\odot$ up to the redshift of 5, in a model-independent framework~\cite{Steinle:2023vxs}. Our model-agnostic results suggest that SMBHs in this range are optimal for \wow{}  detection. Similarly, Ref.~\cite{Sato-Polito:2024lew} argues that nano-Hz results imply that massive BHs may be more abundant than previously expected. Moreover, the James Webb Space Telescope (JWST), telescope has recently observed a large number of massive BHs and galaxies at high redshifts, suggesting higher-than-expected masses of SMBHs and rate of mergers in the LISA band~\cite{kocevski2024risefaintredagn,Harikane_2023}. Overall, this shows that it may be possible to observe \wow{} in LISA, and probe different types of matter acting as gravitational lenses. At the same time, the non-observation can also be used to constrain the abundance of each of these objects till very high redshifts.

\section{Discussion}\label{sec:discuss}

We will now discuss how some of our assumptions affect the probabilities we have computed. Sec.~\ref{sec:discuss_dm} examines how increasing the dark-matter halo concentration dramatically improves the detection prospects. In Sec.~\ref{sec:discuss_SL} we discuss the strong lensing limit of our lensing estimates. And, Sec.~\ref{sec:discuss_symmetry} discusses the potential impact of assuming symmetric and isolated lenses and a comparison of our model assumptions with existing results on the strong lensing optical depth.

\subsection{Role of dark-matter properties}\label{sec:discuss_dm}

Our analysis suggest that MBBHs observed by LISA can probe of dark-matter, particularly for theories that increase the central density of dark-matter halos.
There are two reasons for this.
First, \wow{}  are very sensitive to the central density of the lens. Second, dark-matter halos and subhalos are extremely abundant, making them one of the most important contributions to the optical depth despite being the least compact lenses. 
To assess this potential application, we will now examine how the mass-concentration relation impacts optical depth.

As discussed above, \wow{} are sensitive to halos and subhalos with $10^{6}-10^{8}~M_{\odot}$. All our results assumed a fiducial mass-concentration relation~\cite{Ishiyama:2020vao}. Simulations of lighter halos are compatible with this assumption~\cite{Wang:2019ftp}, but rely on standard assumptions about dark matter and initial conditions. With this relation, lighter halos have higher concentrations (smaller scale/virial radius ratio), but lower convergence in their central area. This lower projected density is ultimately responsible for the lower detection probability.
As an example, we investigate the impact of boosting the mass-concentration relation for light (sub)halos.
%%%%%%%%%%%%%%%%%%%%%%%%%
%%------FIGURE-----------
%%%%%%%%%%%%%%%%%%%%%%%%%

\begin{figure}[t]
    \centering
    \includegraphics[width=0.99\columnwidth]{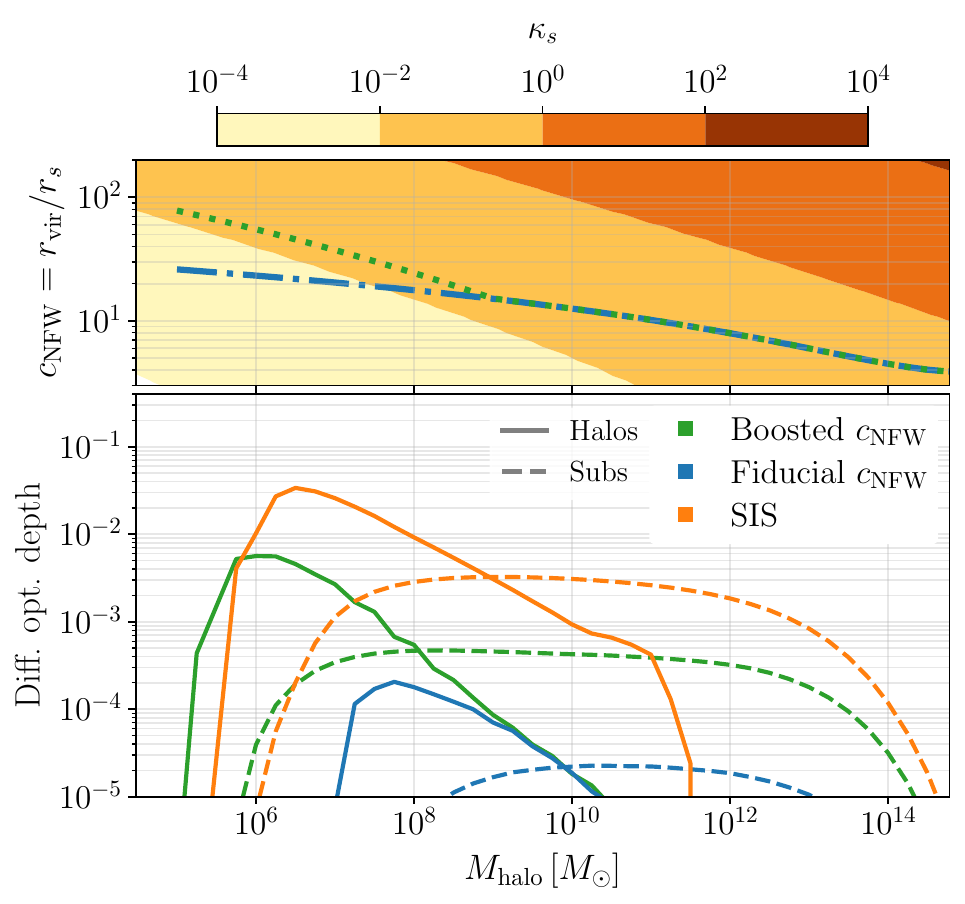}
    \caption{\textbf{Top:} NFW concentration parameter as a function of virial mass for our fiducial model (blue dash-dotted) and the boosted case (green dotted). Colorbar shows the increase in the convergence of the lens at the scale radius. \textbf{Bottom:} Differential optical depth for halos (solid) and subhalos (dashed) modelled as an SIS (orange), NFW with fiducial concentration parameter (blue) and NFW with boosted concentration parameter (green). The source is an equal-mass, quasicircular and non-spinning massive black hole merger with total mass $M_{\rm BBH}=10^{6}~M_{\odot}$ at redshift $z_{S}=5$.}
    \label{fig:discuss_dm}
\end{figure}
%%%%%%%%%%%%%%%%%%%%%%%%%
%%------FIGURE-----------
%%%%%%%%%%%%%%%%%%%%%%%%%

Fig.~\ref{fig:discuss_dm} shows the result of increasing the NFW concentration. The top panel shows the fiducial and boosted mass-concentration relation at $z=0$ (lines), plotted over a colormap of $\kappa_{s}$, the convergence at the scale radius. The modification is given by:
\begin{equation}\label{eq:cnfw_boost}
    c_{\rm NFW} = \left\{ \begin{array}{ll}
         c_{\rm NFW}^{\rm fid.} & \mbox{if $M_{\rm vir} \geq 10^{9}~M_{\odot}$}\\
        3\times c_{\rm NFW}^{\rm fid.} & \mbox{if $M_{\rm vir} < 10^{9}~M_{\odot}$},\end{array} \right.
\end{equation}
which amounts to a factor 3 at $M_{\rm vir}\sim 10^9 M_\odot$ relative to the fiducial model, but with vanishing deviation for $M_{\rm vir}\gtrsim 10^{9}~M_{\odot}$. We have chosen this virial mass cut because it is slightly smaller than the cutoff mass where fitting formulae for the NFW concentration parameter resides~\cite{Diemer:2018vmz, Ishiyama:2020vao}.
The bottom panel shows the corresponding differential optical depth for the LISA source discussed in Sec.~\ref{sec:probs_iso} ($M_{\rm BBH}=10^6,z_{S}=5$) for the fiducial NFW (blue), increased concentration (green) and SIS profile (orange). 
 
More compact (sub)halos have a larger probability of detection, as expected. The total dark matter optical depth for fiducial and boosted NFW profiles in Fig.~\ref{fig:discuss_dm} is, respectively, $4\times 10^{-4}$ and $1.05\times 10^{-2}$, and the SIS profile gives $7.05\times 10^{-2}$. In addition to boosting the overall optical depth, increased concentration allows access to lighter halos. This trend is consistent with the increase in $y_{\rm cr}$ and the approximate location of the $F(w)$ peak for $M_{Lz}y_{\rm cr}^2\sim 1$ (i.e. at $y_{\rm cr}$ corresponding to the Fresnel scale).
The reason for this increase is apparent from Fig.~\ref{fig:diagram_ampl_fact}, where NFW profiles with larger $\kappa_s$ produce stronger \wow{}.
Note that a factor $3$ increase in concentration produces an optical depth 26 times higher. This example highlights the potential of \wow{}  to probe the compactness of (sub)halos and test dark-matter theories.

\subsection{Strong lensing limit}\label{sec:discuss_SL}
%%%%%%%%%%%%%%%%%%%%%%%%%
%%------FIGURE-----------
%%%%%%%%%%%%%%%%%%%%%%%%%
\begin{figure}
    \centering
    \includegraphics[width=0.99\columnwidth]{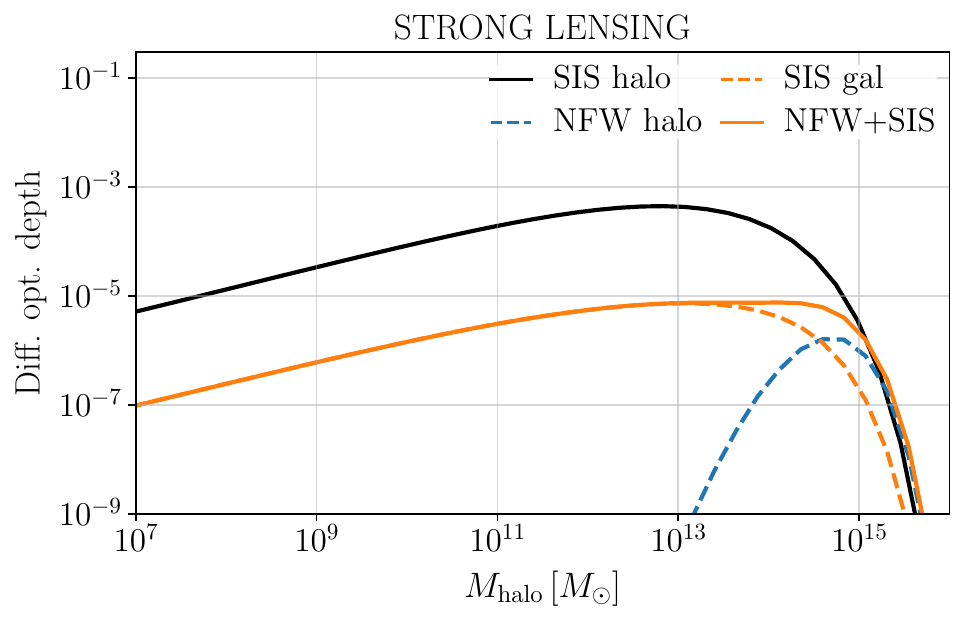}
    \caption{Differential optical depth for strong lensing (multiple images).  The solid black line represents the curve computed assuming DM halos are modelled by SIS. The orange line corresponds to the composite lens case, consisting of NFW halos (dashed blue) and SIS central galaxies (dashed orange). The result is independent of the source mass, assumed at $z_S=5$. }
    \label{fig:opt_depth_SL}
\end{figure}
%%%%%%%%%%%%%%%%%%%%%%%%%
%%------FIGURE-----------
%%%%%%%%%%%%%%%%%%%%%%%%%

To understand the impact of our assumptions, we can compare our results with analyses based on simulations. Although no simulation-based studies of \wow{} exist, we can compare predictions of strong lensing optical depth using our assumptions. A recent study~\cite{Robertson:2020mfh} investigated the strong lensing optical depth using results from two cosmological hydrodynamical simulations, each covering a wide range of massive halos and galaxies. 
The study finds that the optical depth for strong lensing (multiple images and high magnification, $\mu>10$) is approximately described by the halo-mass function, where each halo has the SIS profile (cf. Fig. 5 in~\cite{Robertson:2020mfh}). In contrast, applying our model based on symmetric lenses (halos, subhalos, and their galaxies) leads to much lower values. This is shown in Fig.~\ref{fig:opt_depth_SL}, where the differential optical depths corresponding to our and the SIS-halo models are compared. Integrating these curves gives $\lambda_{\rm SL}^{\rm model}/\lambda_{\rm SL}^{\rm SIS-halo}= 1.8\times 10^{-2},\, 1.1\times 10^{-3}, \, 2.2\times 10^{-2}$ for galaxies, halos and the sum of both (this number is independent of the magnification threshold). 
Our results for the optical depth for multiple-image formation are similar to Ref.~\cite{Tizfahm:2024imk}.

The vast discrepancy could be partly explained by our simplified modelling: strong lensing probabilities are significantly enhanced by collective effects (e.g. extended caustic networks), present in realistic matter distribution analogously to the isolated point lens.  As pointed out by~\cite{Robertson:2020mfh} the probability of strong gravitational lensing from galaxies can be estimated by an SIS profile combined with a usual halo mass function, such as the Tinker fit used in this work, even though galaxies have different mass and number distributions than of their hosts. Like in the strong lensing case, which is mostly sensitive to the mass enclosed within the Einstein radius of the lens, we have found out that \wow{} are considerably sensitive to the inner profile of their lens model. We believe that our simplifying assumptions of symmetric and isolated lenses may underestimate the prospects of detection. Therefore, a full non-perturbative approach using high-resolution cosmological hydrodynamical simulations would considerably help to fully understand all the effects that contribute to \wow{} detection.

\subsection{Symmetric lens analysis} \label{sec:discuss_symmetry}

In this work, we have assumed symmetric lenses and considered them in isolation. Although providing a simple starting point, these simplifications can significantly affect the results. An example in which qualitatively different results are obtained is the point lens: In isolation, it always produces two images, but adding a perturbation causes 4 images to form in an extended region of the source plane. The 2 and 4 image regions are separated by an extended caustic, in which the magnification diverges~\cite{Schneider:1992}. This rich structure is completely missed for symmetric lenses.

Could collective effects lead to a comparable enhancement of \wow{}  detection probabilities? Addressing this question is beyond the scope of this work, but we can use a simple example to investigate the effect of the environment on \wow{}  detection. Let us consider a subhalo in close proximity to a more massive halo, such that $M_{\rm sub}/M_{\rm halo}=0.1$, both described by an SIS profile, and with a separation $x_{\rm sub}=2.5$ (in units of $\xi_0$, as defined by the main halo scale). We will then consider a source that produces a single image at fixed distance from the subhalo $x_{\rm I-sub}=0.4$, and explore the impact of the image's position relative to the halo-subhalo axis. This offers an idea of how the nonsymmetric environment affects the \wow{}  produced by the subhalo.

%%%%%%%%%%%%%%%%%%%%%%%%%
%%------FIGURE-----------
%%%%%%%%%%%%%%%%%%%%%%%%%
\begin{figure*}[t!]
    \centering
    \includegraphics[width=\linewidth]{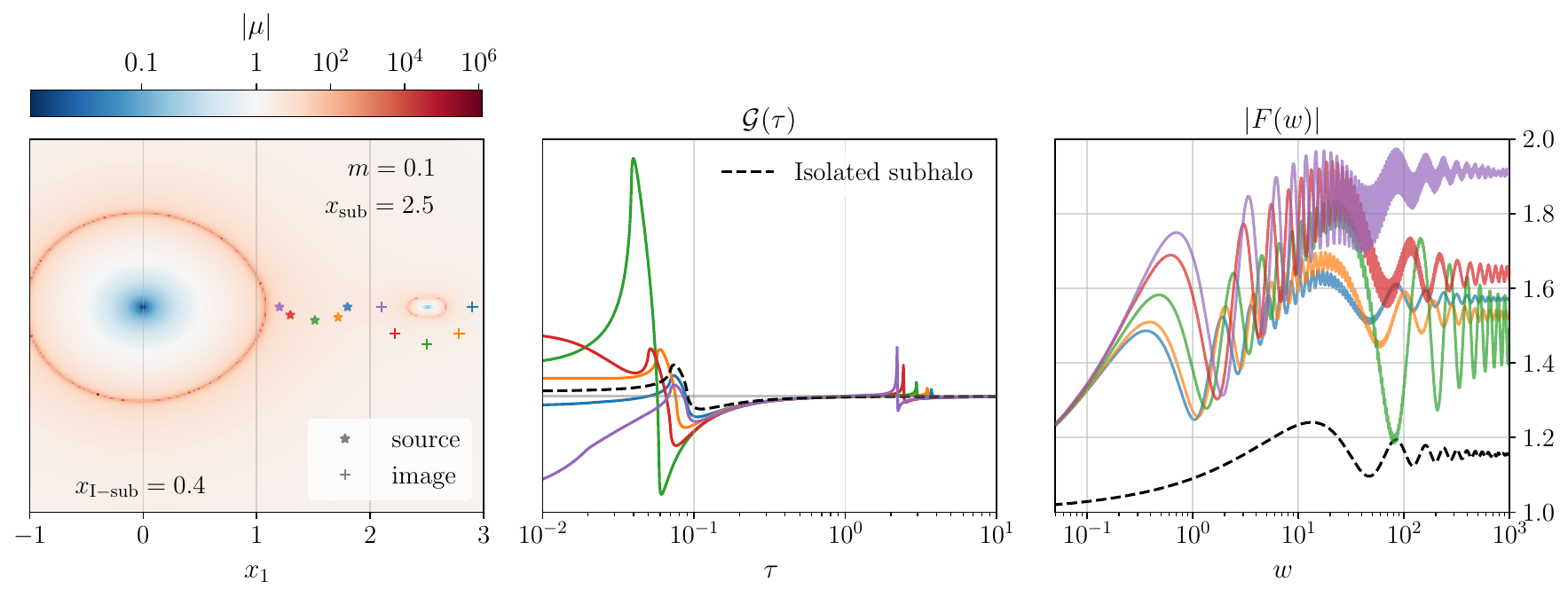}
    \caption{\wow{}  enhancement by collective effects.  We study lensing by the composite potential of a central main halo and a subhalo located at $x_{\rm sub}=2.5$ with $M_{\rm sub}/M_{\rm halo}=0.1$. Both halos are modelled by SIS lenses.
    \textbf{Left:} geometric setup, colors denote the absolute value of the GO magnification ($|\mu|\to 0$ in blue, $|\mu|\to \infty$ in red). Source positions and their corresponding (single) images are shown as stars and crosses, respectively. The images are at a $x_{\rm I - sub}= 0.4$ distance from the subhalo center.
    \textbf{Center:} Green's function for the sources considered. Note that the first/second peak ($\tau \lesssim 0.1\, /\, 3$) correspond to the sub/main halo. The dahsed black line corresponds to the isolated subhalo contribution.
    \textbf{Right:} amplification factor. \wow{} are strongly enhanced by the system whose image lies in the line perpendicular to the halo-subhalo axis, with the smallest GO magnification (green).
    }
    \label{fig:discuss_collective}
\end{figure*}
%%%%%%%%%%%%%%%%%%%%%%%%%
%%------FIGURE-----------
%%%%%%%%%%%%%%%%%%%%%%%%%

Fig.~\ref{fig:discuss_collective} shows the results of the halo-subhalo \wow{} . The left panel summarizes the setup. Color coding shows the magnification, with red lines corresponding to critical curves ($|\mu|\to \infty$).
The source positions and their corresponding (single) images are indicated by stars and crosses, respectively. The middle and left panels show the Green's function and amplification factor for these sources. Note that two \wow{}  are observed, a first one $\tau\lesssim 0.1$ is associated to the subhalo, whose cusp is closest to the image. The second \wow{}  at $\tau\sim 2-5$ is due to the main halo. The two \wow{} in addition to the GO image appear as a secondary modulation in $|F(w)|$. Note that the GO magnification and the \wow{}  amplitude have a very different trend. 

We observe that some configurations provide a marked enhancement of the subhalo's \wow{}. The largest \wow{}  corresponds to the GO image being perpendicular to the halo-subhalo axis, where the peak of the Green's function is boosted with respect to other source positions. 
Note that this case corresponds to the smallest GO magnification. Because $\mu\equiv \det{\frac{d\mathbf{x}}{d \mathbf{y}}}$ encodes the ratio of area in the lens/source planes, this situation is associated to a higher probability.

The comparison with the strong-lensing optical depth and the example of enhanced \wow{} by the higher concentration halos and by the non-axisymmetric lenses suggest that our results are conservative with respect to lens modelling. Future work on this matter should include more realistic matter distribution and an understanding of the effects of non-symmetric lenses. Nevertheless, we also note that other studies have found moderate values of shear or ellipticity to not affect the strong-lensing optical depth significantly~\cite{Huterer:2004jh}.

\section{Conclusions}\label{sec:concl}

In this work, we studied the probability of detecting \wow{} in gravitational waves for LISA sources in the single-image regime. 
This extended our previous analysis~\cite{Savastano:2023spl} by performing realistic lens modelling in gravitational wave lensing, consistently incorporating dark matter-only lenses (halos and subhalos) modelled as NFW, baryonic lenses (galaxies) modelled as SIS, and the presence of a single SMBH modelled as PL in each galaxy. The abundances and masses of lenses reflect available prescriptions in the literature, such as the halo and subhalo mass functions and the stellar-to-host mass relation to determine the mass of each galaxy. This methodology was thoroughly presented in Sec.~\ref{sec:setup}.

To estimate lensing probabilities we have computed the critical impact parameter, $y_{\rm cr}$, in which \wow{} contributions can be detected, given different lens profiles. This was done in~Sec.~\ref{sec:detectability}, where we also discussed the role of degeneracies with source parameters by comparing the usual mismatch criterion, Eq.~\eqref{eq:def_mismatch}, with the fitting factor, Eq.~\eqref{eq:fitting_factor_def}, in which intrinsic parameters are marginalized over. We verified that even though the fitting factor gives more pessimistic detection prospects, the difference between the mismatch criterion and the fitting factor is small for most lens masses, see Fig.~\ref{fig:ff}.

In Sec.~\ref{sec:probs_iso} we computed the differential optical depth for the lenses under consideration. We first quantified the difference between modelling dark matter-only structures, halos and subhalos, using SIS and NFW profiles. We found an overall reduction of two orders of magnitude in the total optical depth of halos and subhalos between SIS and NFW. Our results for both lensing estimates agree with previous results in the literature \cite{Urrutia:2023mtk,Savastano:2023spl} for both cases. Still in Sec.~\ref{sec:probs_iso}, we have considered the contribution to the differential optical depth from galaxies (central and satellites) and their corresponding SMBHs.  Although galaxies are more compact than their hosts, halos and subhalos, their abundance is not significant enough to boost up their chance of detection. Our analysis concludes that SMBHs in the center of galaxies are the major contributors to the \wow{} lensing probability.

In our SMBH modelling, we have considered a fiducial mass relation between SMBHs and their host galaxies of $M_{\rm SMBH}=0.01\times M_{\rm gal}$. Because the precise relation is still a topic of debate~\cite{smbhs_hydro}, in Fig.~\ref{fig:lens_prob_var_smbh} we explored two additional values of the ratio $M_{\rm SMBH}/M_{\rm gal}$, $0.005$ and $0.02$, and found that in the latter case leads to an increase of two times in the optical depth of SMBHs. 
Also, our modelling of galaxies assumes a fixed ratio between the galactic radius, $R_{\rm gal}$, and their host halo or subhalo, $R_{\rm host}$, to $R_{\rm gal} = 0.02 \times R_{\rm host}$. In Fig.~\ref{fig:gals_vary} we showed that reducing the galactic radius by a factor of five with respect to its baseline value of $0.02$ corresponds to an enhancement of twenty times in the total optical depth of galaxies. The variation of the SMBH mass and the size of the galactic radius suggests that the prospect of (not) detecting \wow{} can also shed light on galaxy formation models and the mass of SMBHs.

In Sec. \ref{sec:lisa_cats}, we extend our analysis to the population of binary sources that can be detectable by LISA. We considered both a model-agnostic and an astrophysical source populations from Ref.~\cite{Klein:2015hvg}, and calculated the lensing optical depths for each case. The lensing optical depths are higher for the high-SNR sources, with the greatest contribution coming from SMBHs in most cases. However, for the low mass (and low SNR) sources, we find that the main contribution is from central galaxies. \wow{} detectability depends strongly on the source properties, especially the SNR. Our results for synthetic LISA population indicate a low probability of detecting \wow{}, with comparable numbers across the three formation channels we have considered. We predict from our conservative lens modelling that an average lensing probability of $1.8 \times 10^{-3}$ - $ 5.8 \times 10^{-4}$ and about $\mathcal{O}(0.1)$ - $\mathcal{O}(1)$ MBBHs to have detectable \wow{} in the 5 years of LISA mission,  depending on the population model, as summarized in Tab.~\ref{tab:lisapop}. These rates and probabilities increase by an order of magnitude if we consider halos and subhalos modelled by an SIS lens profile, as anticipated in Sec.~\ref{sec:probs_iso}.

Our modelling of dark matter (sub)halos was conservative. In Sec.~\ref{sec:discuss_dm}, we explore how a boost in the NFW concentration parameter for low-mass halos affects detection rates, a possible outcome of some dark matter theories. In Eq.~\eqref{eq:cnfw_boost}, we multiplied our fiducial model for $c_{\rm NFW}$ from~\cite{Ishiyama:2020vao} by a factor of three for masses smaller than $10^{9}~M_\odot$. 
Increasing the compactness of halos, we further increase the convergence at the scale radius, $\kappa_{s}$, which translates to an enhancement in the optical depth of 26 times, see top and bottom panels in Fig.~\ref{fig:lens_prob_boosted}. This demonstrates how weakly lensed GWs can become a powerful probe of dark matter models in environments, complementary to the strong-lensing regime.

In Section~\ref{sec:discuss_SL}, we tested our model in the strong lensing limit, where recent results from numerical cosmological simulations are available \cite{Robertson:2020mfh}. Our optical depth estimates are 2 orders of magnitude smaller than simulation results. Interestingly, Ref.~\cite{Robertson:2020mfh} found that modelling halos with SIS profile gives a good approximation to the optical depth from simulations (cf. Fig~\ref{fig:opt_depth_SL}). This comparison suggests that our modelling might be too conservative. 
Motivated by this observation, in Sec.~\ref{sec:discuss_symmetry} we consider the role of non-symmetric lenses on \wow{}  detectability. Through a simple example of a two-lens system,   Fig.~\ref{fig:discuss_collective}, we show how collective effects can significantly enhance \wow{}. These considerations indicate that the results produced by our lens model are conservative.

Our main results can be summarized as follows: 
\begin{enumerate}

\item \wow{}  detection is highly sensitive to lens profile. Dense objects are easier to detect, with the point lens as the most favorable (Fig.~\ref{fig:diagram_ampl_fact}). Nonetheless, extended lenses exhibit a larger \wow{} cross-section in the low-mass range, despite being less compact (Fig.~\ref{fig:y_crit_comparison}). Consequently, SMBHs dominate the \wow{} optical depth, followed by galaxies and DM halos and subhalos. The detection probability  within our model ranges from $1.8 \times 10^{-3}$ - $ 5.8 \times 10^{-4}$, see Tab. \ref{tab:lisapop}. 
Modifying our model to recover the correct strong-lensing rates (halos and subhalos as SIS) increases \wow{} detection prospects by an order of magnitude.

\item Increasing the central density of dark-matter (sub)halos boosts the probability substantially: a factor $\sim 3$ higher concentration increases the optical depth by $\sim 26$ (Sec.~\ref{sec:discuss_dm}). This suggests that LISA (non)-observation of \wow{} can probe different dark matter models, as well as probe the mass concentration relation of low mass cold dark matter halos, similarly to~\cite{Gilman:2019bdm}.

\item In addition to central density, a given lens' abundance is also crucial. In this way, the high abundance of halos and subhalos partly compensates for their low (individual) cross section.

\item  Like previous studies, we find that the detection significance criterion can highly impact the \wow{} detection estimates (Table~\ref{tab:halos_subs_gals_bhs}). This is also in agreement with the high correlation between SNR and optical depths found in Sec. \ref{sec:pop_agnostic}. 
On the other hand, considering fitting factors instead of mismatch in cross-section computations lowers the probabilities only slightly, indicating small degeneracies between the source and the lens parameters.

\item Deviations from axial symmetry and collective effects are expected to enhance \wow{}  detectability. We have shown this explicitly with a 2-halo example (Fig.~\ref{fig:discuss_collective}) in which the main halo can boost the \wow{} dramatically.

\end{enumerate}
We now discuss how our results can vary under different model assumptions, before sketching future research directions.

There is the possibility that real-universe lenses are more concentrated than assumed here, which would increase \wow{} detection rates. For example, observations of strong lensing are under-predicted by simulations, a discrepancy that could be reconciled by more concentrated halos~\cite{Meneghetti:2020yif,Meneghetti:2023fug,Dutra:2024qac}. 
A similar argument may apply to galaxies, where observations find a steeper mass-radius relations than SIS~\cite{Oguri:2013mxl,Sonnenfeld2019}. Similarly, the profile of subhalos has been shown to be different than the NFW profile we used, and recent studies~\cite{Heinze:2023ycq} have shown that the NFW does not accurately capture shubhalo profiles, with subhalos having a steeper inner slope in the presence of baryons. In Ref.~\cite{Heinze:2023ycq}, the process of tidal stripping of subhalos plays an important role in changing their concentration, with subhalos closer to the host's center having larger concentration parameter. Given the role of concentration on detection prospects, these effects could increase \wow{} detectability, even if only symmetric and isolated lenses are considered.

Our knowledge of high redshift lenses is limited by the EM observations. In future, studies using surveys from instruments like JWST will definitely help in reducing the uncertainties in the our lens modelling and calculation of probabilities. Recent observations of JWST suggest a larger number and heavier SMBHs exist at high redshifts, relative to previous expectations \cite{kocevski2024risefaintredagn,Harikane_2023}. Observations also indicate that galaxies at the pre-ionisation epochs are more compact and have higher star formation efficiencies \cite{JWSTmorphology}, which could also increase their lensing cross-sections.

GW source properties can also affect the detection prospects. We have considered quasi-circular sources, and included the effects of the high mass ratios simply as a rescaling of the SNR, without accounting for higher-order modes in the mismatch calculations. Including higher-order modes can increase our sensitivities as they will also carry different \wow{} and may lead to higher Bayes' factors, particularly for spin-precessing and eccentric systems~\cite{Urrutia:2024pos}. In the future, we hope to incorporate these effects for more accurate results.  Finally, there may be more massive black hole binary mergers than expected in the LISA frequency band. This possibility remains compatible with nanohertz GW observations from pulsar-timing arrays~\cite{Steinle:2023vxs,Sato-Polito:2023gym,Sato-Polito:2024lew} and JWST observations as discussed above.

Our investigation indicates future promising directions on \wow{} for low-frequency GW detectors. Potential applications include the identification of \wow{} in the Bayesian model selection framework. To correctly identify the lens and measure its properties one needs to perform parameter estimation runs using each lens model,  currently, this is possible only for stellar-mass binaries assuming symmetric lens models (e.g. SIS, NFW, and PL) \cite{Wright_2022,Cheung:2024ugg} and for overlapping multiple lensed images in the GO limit \cite{Liu_2023,Toscani:2023gdf}. These runs for massive binaries are computationally expensive and recent studies have proposed to speed it up by using relative binning and GPUs~\cite{weaving2023adapting,Hoy:2023ndx,Karnesis_2023}. Therefore, detecting \wow{} in LISA in a Bayesian framework is a challenge and requires further work. It is also worth mentioning that if not accounted for, lensing can bias various tests of GR and measurement of spins \cite{10.1093/mnras/stae836,Mishra:2023vzo}.

The possibility of \wow{}  from objects with different density profiles raises the question of potentially distinguishing them using data. A distinction is possible in principle~\cite{Cremonese:2021ahz}, and it may even allow a reconstruction of the density profile under symmetry assumptions~\cite[Sec.~VA]{Savastano:2023spl}. However, a practical implementation of lens model selection must be performed. 
For the case of LISA, the role of lensing signatures should be addressed in the context of a global fit analysis.
The prospect of \wow{}  detection is likely to improve if other proposals for GW detection are implemented~\cite{Sesana:2019vho,Sedda:2019uro,Baibhav:2019rsa,Gong:2021gvw,Ajith:2024mie,Zwick:2024hag}.

A clear application of \wow{}  is to test the properties of dark matter. The steep dependence of the \wow{}  optical depth with the halo concentration (Fig.~\ref{fig:discuss_dm}) suggests that LISA will be able to test theories where $10^6-10^9M_\odot$ (sub)halos are more concentrated than in $\Lambda$CDM. This possibility remains true even when modelling lenses as isolated. Primordial black holes~\cite{LISACosmologyWorkingGroup:2023njw} are highly constrained in the window that can be probed by \wow{}~\cite{Serpico:2020ehh,Agius:2024ecw}. However, \wow{}  can be used to test other theories that produce compact structures~\cite{Zumalacarregui:2024ocb,Graham:2024hah,Croon:2024rmw}. These include ultralight fields~\cite{Ferreira:2020fam,Hui:2021tkt} (via attractive self-interactions of the field~\cite{Arvanitaki:2019rax}), self-interacting dark matter~\cite{Tulin:2017ara,Adhikari:2022sbh} (via runaway collapse of subhalos~\cite{Gilman:2021sdr,Zeng:2023fnj}) or warm dark matter (via prompt collapse of the first structures~\cite{Delos:2023exh}). 

These investigations of dark matter and baryonic properties will ultimately need to include the effects of non-symmetric matter distributions, as argued above. To this end, results from cosmological simulations need to be combined with insights on light subhalos and central densities, able to produce \wow{}  in the LISA band. A detection of \wow{}  (or its absence) will inform our understanding of astrophysics and fundamental physics in the high-redshift universe. 
Ultimately, we expect that the rich phenomena associated to \wow{} will lead to yet unforeseen applications.

\acknowledgements{
We are very grateful to  M. \c{C}al\i{}\c{s}kan,  M. Cheung, T. Collett, S. Delos, J.M. Ezquiaga, N. Menadeo, G. Tambalo for discussions.
GB is supported by the Alexander von Humboldt Foundation. HVR is supported by the Spanish Ministry of Universities through a Margarita Salas Fellowship, with funding from the European Union under the NextGenerationEU programme. 

The numerical calculations reported in the paper are performed on the Alice computing cluster at ICTS-TIFR, with the aid of \glow{}, \texttt{PyCBC}~\cite{alex_nitz_2023_7885796} and \texttt{Colossus} \cite{Diemer:2017bwl} software packages.
}

\appendix

\section{Lens model details}\label{app:lens_models}
In this Appendix, we introduce the SIS and NFW lens models that are used in the main text, showing their characteristic properties and equations. Notice that what discussed below for halos applies to subhalos too.

\paragraph{Singular Isothermal Sphere}

As a function of the radial coordinate, the SIS profile reads as
\begin{equation}
  \rho(r)  
  =  
  \frac{\sigma_v^2}{2\pi G r^2}  
  \;,
\end{equation}
where $\sigma_v$ represents the velocity dispersion of the halo. The corresponding lensing potential is $\psi(x) = \psi_0 |x|$, with $\psi_0= \sigma_v^2 / (\xi_0 G \, \Sigma_{\rm cr})$. Then, a convenient choice of the normalization scale is $\xi_0 \equiv \sqrt{4G \Mlz d_{\rm eff}} \equiv \sigma_v^2 / (G \, \Sigma_{\rm cr})$, so that $\psi_0=1$ and $\psi(x) = |x|$. 

Considering the SIS to model a galaxy with mass $M_{\rm gal}$ enclosed in the galactic radius, $R_{\rm gal}$, integrating the profile we get
\begin{equation}\label{eq:M_SIS}
   M_{\rm gal}= \frac{2 \sigma_v^2}{G} R_{\rm gal}\,,
\end{equation}
and using $\xi_0$ definition,
\begin{equation}\label{eq:M_lz_M_SIS}
    \Mlz=\left(\frac{1}{2}\frac{M_{\rm gal}}{R_{\rm gal}}\right)^2 \frac{\pi (1+z_L)}{\Sigma_{\rm cr}}\,.
\end{equation}
Alternatively, when the SIS is used to model DM halos of a certain virial mass, $M_{\rm halo}$, enclosed a virial radius defined by $\rho(R_{\rm vir})\equiv 200 \rho_c$, we obtain \cite{Savastano:2023spl}
\begin{align}\label{eq:MlzMvir_SIS}
    \Mlz
    &=
    \frac{4 \pi^2}{G}
    (1+z_L)^2 d_{\rm eff}
    \left(
    \frac{5\sqrt{6}}{2}G H(z_L) M_{\rm halo}
    \right)^{4/3}
    \\ \nonumber
    &=
    2.3
    \times 
    10^6 \, M_\odot (1 + z_L)^2
    \left(
    \frac{d_{\rm eff}}{1\,{\rm Gpc}}
    \right)
    \times
    \\ \nonumber
    &
    \hspace{3cm}
    \times
    \left(
    \frac{M_{\rm halo}}{10^{9} \, M_\odot} \frac{H(z_L)}{H_0}
    \right)^{4/3}
    \,.
\end{align}

In the GO regime, the SIS profile produces two images, for any lens mass, if the impact parameter is inside the caustic line $y_{\rm sl}=1$. They are a a minimum and a saddle point of the Fermat potential, (labelled respectively by "+" and "-"), and have $\mu_{\pm} = 1/y\pm 1$, time delays $\phi_{\pm} = \mp y -1/2$ and Morse phases $n_+ = 0$, $n_- = 1/2$. In weak lensing, at $y>\ycr{}$ only one image survives.

\paragraph{Navarro-Frenk-White}

One of the most commonly used profile to fit dark matter halos is the 
the Navarro-Frenk-White (NFW), which scales with the distance $r$ as
\begin{equation}\label{eq:nfw_rho}
    \rho(r)    =    \frac{\rho_s}{(r/r_s) (1 + r / r_s)^2}
    \;,
\end{equation}
where $r_s$ is the scale radius and $\rho_s/4$ is the density at $r_s$. 
If we define $u \equiv x/x_s$ and $x_s=r_s/\xi_0$, the convergence is
\begin{equation}
\kappa(x)= 2\kappa_s\frac{1-\mathcal{F}(u)}{u^2-1}
\end{equation}
with 
\begin{equation}
\kappa_s= \frac{\rho_s r_s}{\Sigma_{\rm cr}}\,\quad{\rm and}\quad
\mathcal{F}(u)=\begin{cases}\frac{\tanh^{-1} \sqrt{1-u^2}}{\sqrt{1-u^2}}&u<1\\1&u=1\\\frac{\tan^{-1} \sqrt{u^2-1}}{\sqrt{u^2-1}}&u>1\end{cases}\,.
\end{equation}
Then, lensing potential follows from the 2D Poisson equation as
\begin{equation}\label{eq:nfw_psi}
    \psi(x)=\frac{\psi_0}{2}\left[\ln^2\frac u2+(u^2-1)\mathcal{F}^2(u)\right]\,,
\end{equation}
with
\begin{equation}
    \psi_0=\frac{4 \kappa_s r_s^2}{\xi_0^2}\,\quad {\rm and} \quad\kappa_s= \frac{\rho_s r_s}{\Sigma_{\rm cr}}\,.
\end{equation}
Introducing $M_{\rm NFW}=4\pi \rho_s r_s^3$ and requiring $\psi_0=1$, the normalization scale reads $\xi_0\equiv\sqrt{4 \pi d_{\rm eff} \Mlz }$, with $\Mlz=(1+z_L) M_{\rm NFW}$.

We now connect the lens parameters to the halo physical parameters. First, we introduce the concentration parameter 
\begin{equation}
    c_{\rm NFW}\equiv\frac{r_{\rm vir}}{r_s}\,,
\end{equation}
with $r_{\rm vir}$ the virial radius of the halo. This quantity can be inferred from numerical simulation or analytical arguments \cite{Correa:2015dva}, and it generally depends on the virial mass and the redshift of the lens. In this work, we adopt as fiducial the model developed in Ref. \cite{Ishiyama:2020vao}, implemented in the python toolkit \texttt{colossus} \cite{Diemer:2017bwl}.  We defined the virial radius following Ref. \cite{Diemer:2018vmz} as 
\begin{equation}
   r_{\rm vir} \equiv\left(\frac{4}{3} \frac{M_{\rm halo}}{ \pi \Delta\rho_c }\right)^{1/3}
\end{equation}
where $\rho_c$ is the critical density of the Universe and the overdensity parameter is set to $\Delta=200$. Because $M_{\rm halo}$ is by definition the mass contained inside the virial radius, integrating the density profile in Eq.~\eqref{eq:nfw_rho} up to $r_{\rm vir}$, we derive the relation between lens and virial mass:
\begin{equation}\label{eq:MlzMvir_NFW}
    M_{\rm halo}=\left[\log{(1+c_{\rm NFW})-\frac{c_{\rm NFW}}{1+c_{\rm NFW}}}\right]\frac{\Mlz}{1+z_L}\,.
\end{equation}
At $z_L=2.5$, for the fiducial concentration relation assumed in our analysis, the lens mass differs by a factor of few.
Thus, the NFW halo profile is determined by $M_{\rm halo}, z_L$, and all the other quantities can be inferred given the assumptions above.

\begin{figure}[h]
    \centering
    \includegraphics[width=\columnwidth]{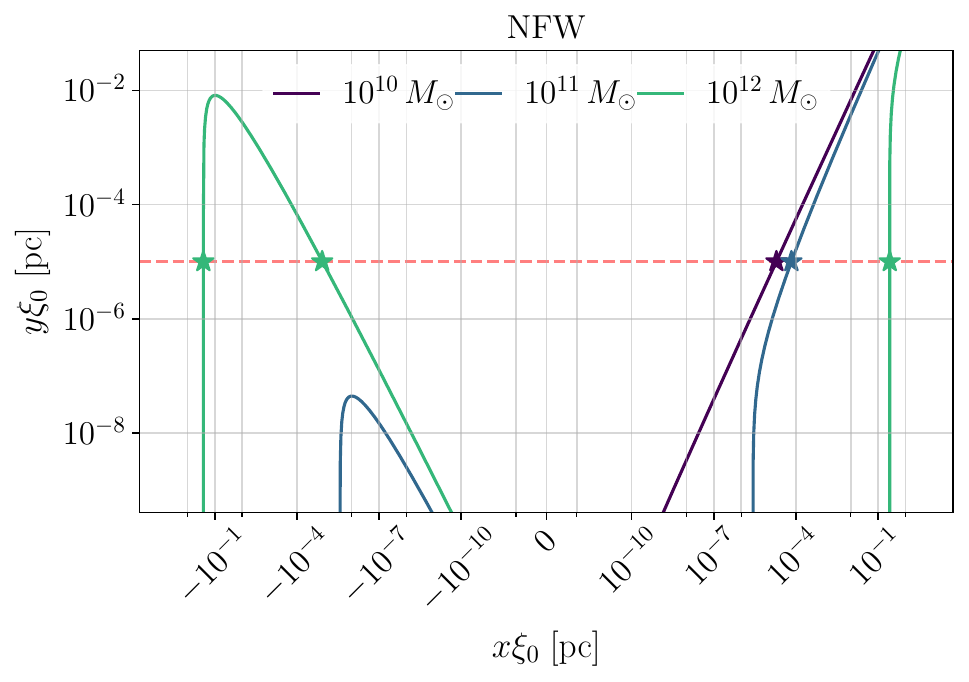}
    \caption{Lens equation for NFW. For a chosen impact parameter, $y$, the image positions are the intersection points (stars) between the dashed red horizontal line and the solid line. Both axis are in physical units.}
    \label{fig:lens_eq_NFW}
\end{figure}

We now discuss the GO limit. Similar to the SIS, the NFW profile also diverges at the origin but with a shallower density profile. As expected for all axisymmetric and \emph{supercritical} lenses ($\kappa(0)>0$), multiple GO images are produced at small impact parameters \cite{Schneider:1992}. However, because the NFW scales as $r^{-1}$ towards the center (compared to $r^{-2}$ of the SIS), it sources 3 images. The image positions on the lens plane can be found numerically by solving the lens equation $\nabla_{\vect x} \phi =0$, as shown in Fig.~\ref{fig:lens_eq_NFW}. In particular, the NFW shows a highly demagnified image close to the lens centre that corresponds to a local maximum of the Fermat potential.

\section{Individual optical depths} \label{app:lensingprob}
\begin{figure*}[tbh]
\centering

\subfigure[ Halos ]{\label{fig:a}
\includegraphics[width=0.45\linewidth]{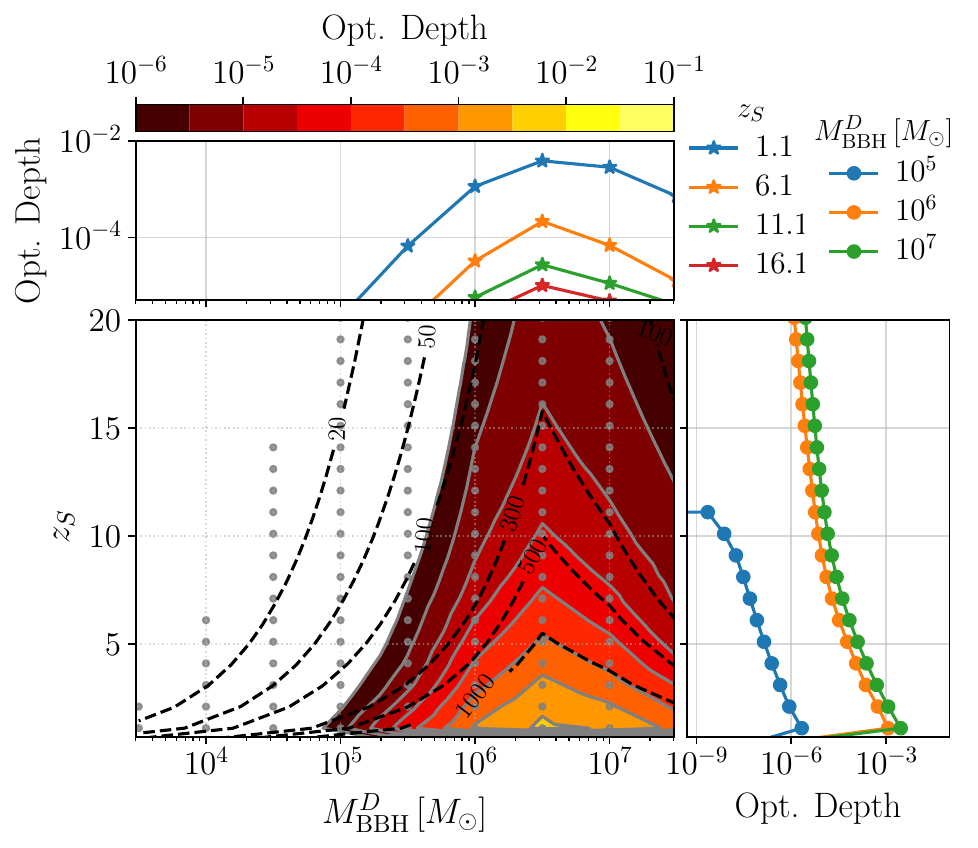}}
\subfigure[ Subhalos ]{\label{fig:b}
\includegraphics[width=0.45\linewidth]{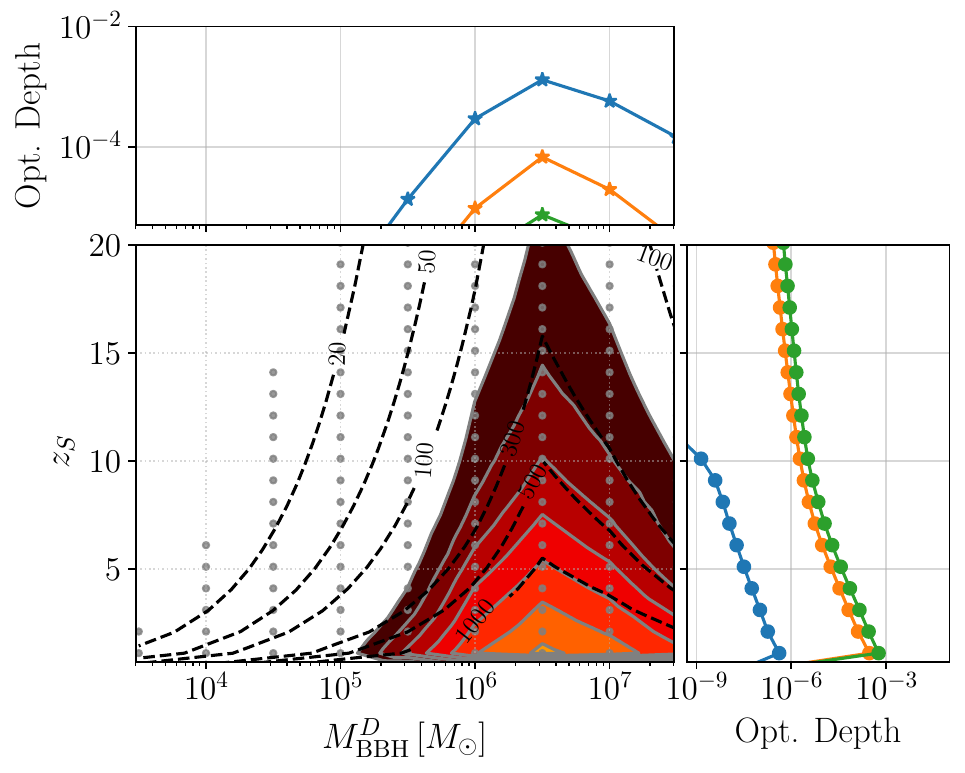}}
\subfigure[ Central Galaxies ]{\label{fig:c}
\includegraphics[width=0.45\linewidth]{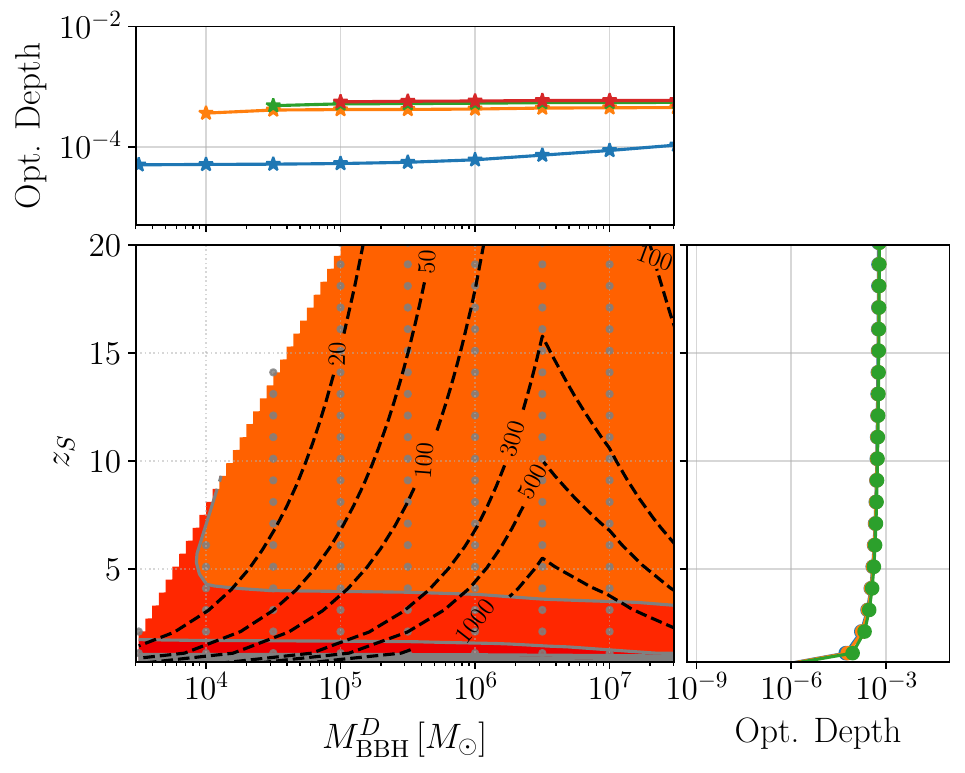}}
\subfigure[ Satellite Galaxies ]{\label{fig:d}
\includegraphics[width=0.45\linewidth]{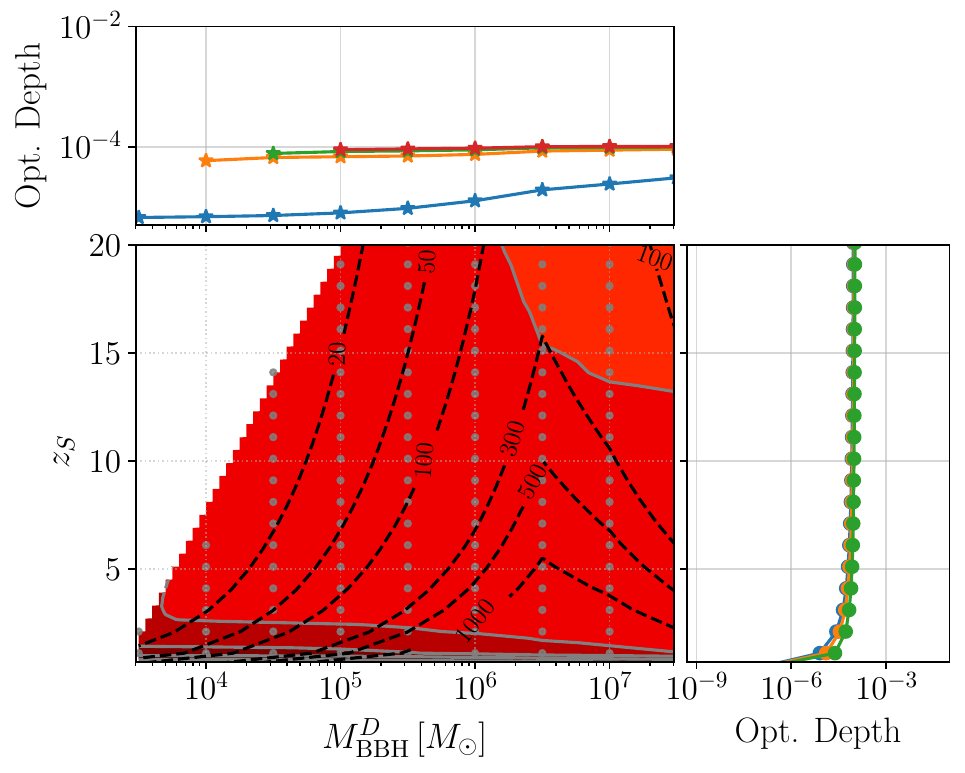}}
\subfigure[ Central SMBHs ]{\label{fig:e}
\includegraphics[width=0.45\linewidth]{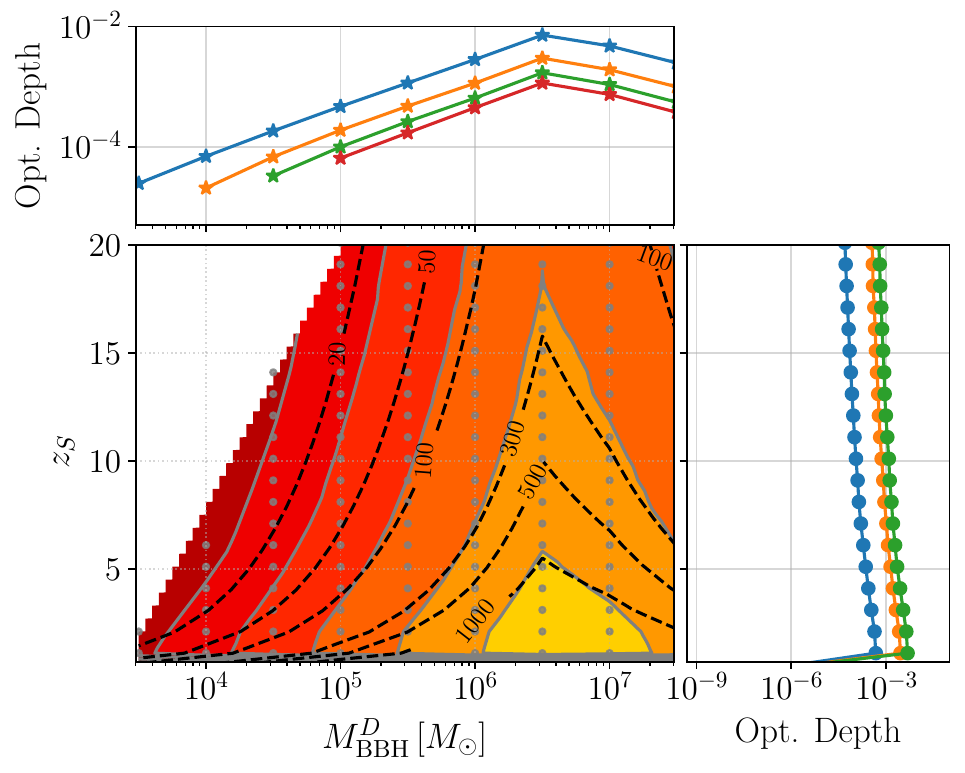}}
\subfigure[ Satellite SMBHs ]{\label{fig:f}
\includegraphics[width=0.45\linewidth]{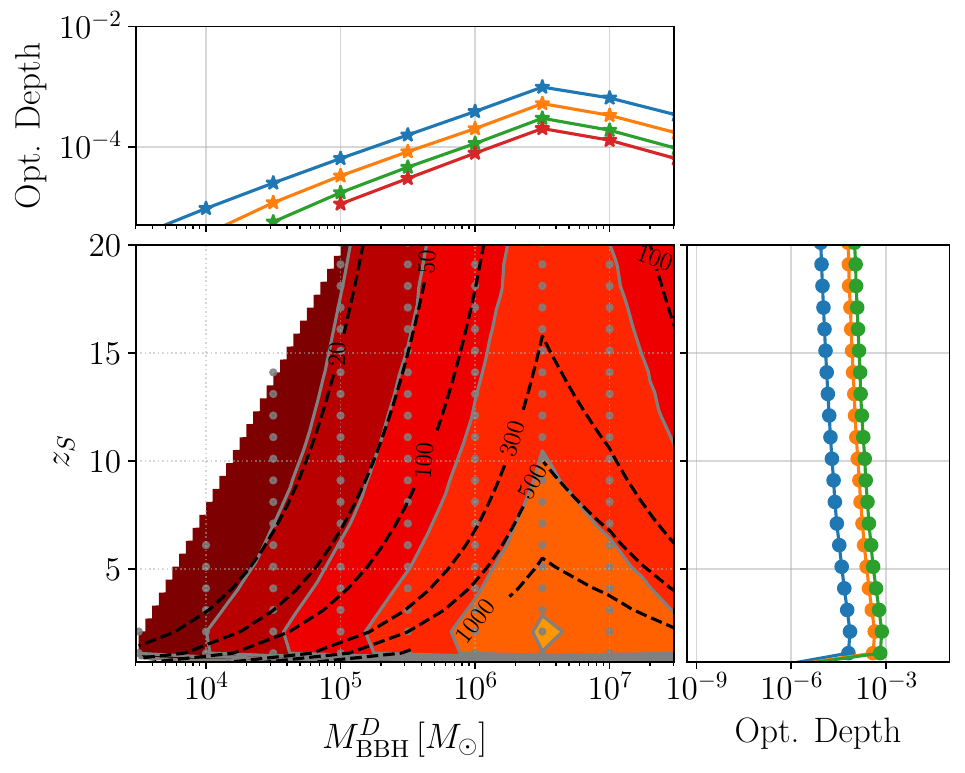}}
\caption{Contribution of individual lenses to the total lensing probability as the function of MBBH source parameters. Black dashed lines are the SNR contours, and gray solid is lensing probabilities.  Except for the galaxy lenses, all types of lenses show a strong correlation with the SNRs. }
\label{fig:iso_lenses_lisa_agnostic}
\end{figure*}

We have modelled the total lensing optical depths as the sum of individual isolated lenses. These individual optical depths vary with the source properties, depending on the source SNRs and mismatch between lensed and unlensed waveforms.  Fig.~\ref{fig:iso_lenses_lisa_agnostic} shows the lensing optical depth as a function of source mass and redshift for different types of lenses. The trend is similar for the lenses with the same mass profile. For halos, subhalos and SMBHs the optical depths contours follow the SNR contours which peak around $M_{\mathrm BBH}^D \sim 5 \times 10^6 M_{\odot}$. On the other hand for galaxies the optical depths are less sensitive to the source properties and for lower total mass MBBHs they get larger than the other types of lenses.

\newpage
\bibliographystyle{apsrev4-1}
\bibliography{gw_lensing}
\end{document}